\documentclass[a4paper,11pt,final,fleqn,epsfig]{article}

\usepackage{natbib}
\usepackage{color, colortbl}
\usepackage{setspace}
\usepackage{bm}
\usepackage{amssymb}
\usepackage{graphicx}
\usepackage{pdflscape}
\usepackage{lscape}
\usepackage{xcolor}
\usepackage{todonotes}
\definecolor{darkblue}{rgb}{0,0,0.5}
\definecolor{Gray}{gray}{0.9}
\usepackage{pstricks,pst-plot}
\usepackage{booktabs}
\usepackage{threeparttable}
\usepackage{todonotes}
\usepackage{subfloat}
\usepackage{subfig}
\usepackage{amsmath,amssymb,latexsym,tabularx}
\usepackage[pdfborder=0, colorlinks=true, linkcolor=darkblue, citecolor=darkblue, urlcolor=darkblue]{hyperref}

\usepackage[ansinew]{inputenc}

\usepackage{courier}

\usepackage[format=hang]{caption} %

\PassOptionsToPackage{gray}{xcolor} %
\usepackage{etex} %
\usepackage{tikz, pgf}

\usepackage[english]{babel}
\usepackage{blindtext}

\begin{document}
\begin{titlepage}
	\title{\scshape{\Large{The Economic Consequences of Being Widowed by War:\\A Life-cycle Perspective}}
		\thanks{We would like to thank seminar and workshop participants at the 2024 annual conference of the Economic History Society, IZA, the IZA workshop on Labor Markets and Innovation during Times of War and Reconstruction, Trinity College Dublin, and the University of Bayreuth for their helpful comments and suggestions. John Tyler Ellis, Bastian Prinz, Charlotte Rehling, and Luis Schr\"{o}der provided excellent research assistance. Any remaining errors are our own. Funding from the Stiftung Bildung und Wissenschaft (project number S0026/10280/2024), MICIU/AEI (grants CEX2021-001181-M and RYC2019-027614-I) and from the APC: Universidad Carlos III de Madrid (Agreement CRUE-Madro\~{n}o 2024) is gratefully acknowledged. This manuscript has been published in the Journal of Public Economics, https://doi.org/10.1016/j.jpubeco.2024.105241, and is made available under the CC-BY-NC-ND license.}}
	
			\author{Sebastian T. Braun\footnote{\small University of Bayreuth, CReAM, and IZA (\mbox{sebastian.braun@uni-bayreuth.de})}
				\hspace*{1.0cm} Jan Stuhler\footnote{\small Universidad Carlos III de Madrid, CReAM, and IZA (\mbox{jan.stuhler@uc3m.es})}} %
			
			\date{\today}
			\maketitle
			\centering
			\vspace{-1cm}
			\thispagestyle{empty}
			\begin{abstract}
				\begin{singlespace}
					Despite millions of war widows worldwide, little is known about the economic consequences of being widowed by war. We use life history data from West Germany to show that war widowhood increased women's employment immediately after World War II but led to lower employment rates later in life. War widows, therefore, carried a double burden of employment and childcare while their children were young but left the workforce when their children reached adulthood. We show that the design of compensation policies likely explains this counterintuitive life-cycle pattern and examine potential spillovers to the next generation.
				\end{singlespace}

				\begin{itemize}
					\begin{singlespace}
						\item[] \textbf{JEL Code}: J16, J20, N34 
						\item[] \textbf{Keywords}: War widows; labor market careers; female labor force participation; World War II
					\end{singlespace}  
				\end{itemize}
				
			\end{abstract}
			
			\thispagestyle{empty}
		\end{titlepage}

		\newpage
\section{Introduction}

In many conflict zones, war widows constitute a large share of the population \citep{Buvinic2012}. For example, World War I (WWI) resulted in 3-4 million war widows \citep{Bette2015}, the Rwandan genocide in 500,000 \citep{UniNations2001}, and one to two million war widows lived in Afghanistan in the mid-2000s \citep{CIRBC2007,Chandran2020}. These women, many with young children, play a central role in rebuilding postwar societies, but we lack evidence on their economic situation \citep{Brueck2009,Brouneus2023}. While there is some work on the (gendered) consequences of spousal loss due to natural causes %
\citep[e.g.,][]{Dreze1997,Burkhauser2005,Fadlon2021}, war widows face unique challenges due to the sudden and violent nature of their spouse's death, their often young age, and limited childcare and financial support in postwar societies.

This paper provides first quantitative evidence on the individual economic effects of being widowed by war. We focus on World War II (WWII), the deadliest conflict in history. Although more than 20 million soldiers died, and the support of their survivors was a key policy issue after 1945, the literature on the individual effects of WWII \citep[e.g.,][]{Ichino2004,Kesternich2014,AkbulutYuksel2022} has not yet examined the economic plight of surviving widows. %
Drawing on rich life course data for West Germany, our analysis focuses on labor market effects over the life cycle. In particular, we test the hypothesis that war widows increased their labor supply due to the adverse income shock of losing their husbands \citep{Boehnke2022}. In addition, we examine whether the daughters of widows held more progressive norms about female employment and increased their labor supply accordingly \citep{Fernandez2004,Gay2023}.

We indeed find that war widowhood significantly increased employment in the immediate postwar period. In 1950, war widows born in 1906-1914 were 13.8 percentage points (pp), or 67\%, more likely to be in market work than otherwise comparable women who did not lose their husbands in the war. However, this positive employment effect gradually declines, and by 1971 war widows were 1.9 pp \textit{less} likely to work (and 2.5 times more likely to rely on welfare) than their peers. This finding %
is especially surprising given that most war widows remained unmarried (they were 73\% less likely to be married in 1971) and labor force participation was higher for unmarried than married women at the time.%

The children of war widows left school and entered the labor force earlier than their peers, presumably due to financial hardship in the postwar period. This educational penalty is particularly pronounced for boys, who lose a full year of education. However, we find no long-run effects of widowhood on either sons' or daughters' employment rates. Overall, widowhood leaves a clear mark on the employment trajectories and socioeconomic outcomes of the 1.0-1.2 million war widows living in West Germany after 1945 \citep{Niehuss2002}, but did not have a lasting effect on women's labor force participation.

We derive these results by comparing war widows, which we can directly identify in our data, with other women who had comparable prewar socioeconomic characteristics. %
Several observations support a causal interpretation of our estimates. First, differences in prewar characteristics between widows and non-widows are small and explain only about 1\% of the variation in widowhood status. %
Second, adding control variables leaves our estimates unchanged while improving model fit. %
Third, %
war widowhood is uncorrelated with spousal or parental characteristics either. These findings are consistent with the fact that the women in our sample were married to men whose cohorts were largely or entirely conscripted for the war, minimizing selection into war service and death \citep{Braun2023}.

The widow's counterintuitive labor supply trajectories are likely explained by the design of the compensation system. Initially, financial support for war widows was limited, forcing many into employment. %
With Germany's economic boom in the 1950s and 1960s, compensation became more generous, especially its means-tested component, which created disincentives to work through standard substitution effects. This pattern is typical: compensation for war widows is often inadequate in war-torn societies, while more generous policies become feasible as the economy recovers and the number of widows eligible for compensation shrinks. \cite{Boehnke2022} document that pension payments to French war widows after WWI were initially low but increased in the 1930s. Similarly, \cite{Skocpol1993} documents how benefits to US Civil War veterans and their dependents expanded massively over time. We find that such back-loaded compensation schemes force widows to bear the double burden of employment and child care while they are young, but incentivize their withdrawal from the labor market later in life.%

Moreover, we show that war-induced labor market entry does not necessarily increase women's labor force participation in the long run. The work experiences of war widows were often negative: working widows were accused of neglecting their childcare responsibilities, and many widows struggled to balance work and household responsibilities in the face of social stigma and a lack of childcare \citep{Schnaedelbach2007}. Likely as a consequence, war widows did not hold more progressive views on the compatibility of family and work, as we show using survey data on attitudes and gender norms. They also did not place a higher value on work than non-widows, likely contributing to their withdrawal from the labor market later in life. In fact, daughters of war widows were more likely to agree with the statement that women with young children should \textit{not} work.

\paragraph{Literature.} Our paper is the first to provide representative individual-level evidence on the effects of war widowhood on labor market outcomes.\footnote{\cite{Salisbury2017} studies the impact of pension payments on the remarriage rates of Union Army widows whose husbands died in the US Civil War.} %
We contribute to a growing literature examining the effects of war mobilization and military fatalities on female labor force participation (FLFP).\footnote{A related literature explores the effect of war-related scarcity of men on the marriage market and fertility \citep{Abramitzky2011,Bethmann2013,Brainerd2017,Kesternich2020,Battistin2022}. \cite{Brodeur2022} connects these two literatures. A more distant literature strand studies the effects of direct exposure to warfare, including our related paper on the individual consequences of battlefield injuries, imprisonment, and displacement \citep{Braun2023}.} %
Much of this literature has focused on the case of WWII in the US, suggesting that the mobilization of men drew women into the labor force \citep{Goldin1991,Acemoglu2004,Goldin2013,Jaworski2014,Doepke2015}. However, recent evidence cautions that WWII decreased female opportunities in white-collar jobs \citep{Bellou2016} and suggests that war production rather than mobilization was the primary cause of the wartime surge in FLFP \citep{Rose2018}. 

While the extant literature emphasizes shifts in aggregate labor demand, we examine how the individual experience of being widowed by war affects women's labor supply over the life cycle. Especially in countries with high military casualty rates, such as Japan, Germany, or the Soviet Union, the experience of war widows is indispensable for understanding the overall impact of WWII on FLFP. Related to our study, \cite{Boehnke2022} note that the negative income shock of losing one's spouse may have led war widows to enter the labor force after WWI. Using variation across French regions, they show that higher military deaths increased FLFP and estimate that widowed women accounted for nearly half of this effect. However, they do not study individual labor market careers%
, as we do here.%

Our paper also relates to studies that analyze the intergenerational persistence of war shocks on FLFP.\footnote{Relatedly, \cite{Dupraz2023} show that sons whose fathers died in the US Civil War had lower incomes later in life, but do not examine the effect on the spouses of fallen soldiers.} %
\cite{Fernandez2004} use variation in WWII mobilization rates across US states to document intergenerational spillovers in women's labor supply. \cite{Fernandez2013} develops a learning model of cultural change, in which WWII mobilization increases FLFP in the next generation, as women drawn into the labor force during the war learn the true cost of working. %
And \cite{Gay2023} finds that women born in French counties that experienced higher military death rates in WWI were more likely to work decades after the war ended. While these studies also capture intergenerational spillovers from demand-driven shifts in women's employment (due to war mobilization), we focus on spillovers from individual changes in labor supply (due to spousal loss).

\section{Institutional Background}
\label{Section Background}

The total number of German military dead and missing was over five million, and a similar number of soldiers were injured \citep{Overmans1999,Mueller2016}. The fate of surviving family members and the war-disabled was thus a pressing social problem in postwar Germany. %
The \textit{Bundesversorgungsgesetz} (BVG) of October 1950 %
aimed at the physical and vocational rehabilitation of war victims and their families, and paid social assistance to those for whom rehabilitation was not, or only partially, possible \citep{Diehl1985}. 

However, the difficult financial situation at the time limited the generosity of the compensation payments. War widows initially received an unconditional basic pension (\textit{Grundrente}) of DM 40 and a means-tested compensatory pension (\textit{Ausgleichsrente}) of up to DM 50 per month. The latter was not paid to women under age 50 unless they had children to care for or were unable to work. The maximum pension of 90 DM represented roughly 30\% of the average gross labor income at the time. Children under age 18 received additional orphan's pensions.

As in other postwar societies \citep{Skocpol1993,Boehnke2022}, pension levels gradually increased over time. By 1960, the maximum amount of basic and compensatory pensions %
was about 49\% of average labor income. Moreover, the conditions for receiving a compensatory pension were lowered and the age limit decreased from 50 to 45 years. %
Additional damage compensation (\textit{Schadensausgleich}) was introduced in 1964 for widows whose income was less than half of what their husbands would have earned, further increasing the maximum pension to nearly 60\% of average income in 1970.
Appendix \ref{sec:compensation} describes how maximum pension payments increased since the mid-1950s, both in real terms and as a share of gross labor income.

Three features of the BVG are worth highlighting. First, the compensatory pension was reduced in proportion to the earned income above a basic allowance. This means-tested part of the widow's pension gained importance from the mid-1950s onwards, especially with the introduction of the likewise means-tested damage compensation (see Appendix \ref{sec:compensation}). Second, women were not entitled to a compensatory pension until later in life, unless they were unable to work or had to care for children. Third, war widows lost their pension rights if they remarried and received a lump-sum settlement upon remarriage.
 
War widows frequently faced a dilemma in their labor market participation \citep{Schnaedelbach2007}. Women's employment was still controversial in the early 1950s, and the employment of war widows was considered particularly undesirable due to concerns they would neglect the care of their children. With scarce institutional childcare available, working mothers often depended on relatives for childcare \citep{Niehuss2002}. %
On the other hand, gainful employment was often essential for financial reasons, given the initially low compensation payments. %

\section{Data and Empirical Strategy}
\label{Section Data}

\subsection{Data}

\paragraph{Microcensus 1971 (MZU71).}
The main data source for our analysis is the MZU71 \citep{Gesis1971}. This representative and mandatory survey provides detailed information on changes in the social and occupational structure of the West German population between 1939 and 1971. Covering 1\% of the population aged 15 and over with German citizenship,  the survey contains information on 456,000 individuals.%

The survey asked respondents in 1971 about their employment status and occupation in 1939, 1950, 1960, and 1971. It also recorded whether respondents owned a house in 1939 and 1971, their education level, their main source of income, and their net monthly income in 1971. The latter is recorded in seven categories and is missing for farmers. %
The survey also contains the place of residence in 1939, which allows us to identify persons displaced from Eastern Europe in the wake of WWII and refugees from the GDR. %

Importantly for our purposes, the survey asked women whether they ``were or had been war widows''. This allows us to identify war widows even if they have remarried. Our treatment group consists of all women who were married at the time of WWII and whose husbands were killed in the war, died in captivity, or were missing in action. We focus on women born in 1906-14 who were between 25 and 65 years old during our observation period (1939-71). Women of the same birth cohort who were married by 1945 but did not lose their spouses in the war form our control group. We drop women who never married or who married only after 1945, as they were not 'at risk' of losing their spouse in the war.\footnote{The MZU71 reports only the year of the last marriage for those married in 1971. Appendix \ref{sec:altcontrol} addresses potential issues from this limitation and presents robustness checks confirming our main findings with alternative control groups. Additionally, we demonstrate that our results hold in the German Life History Study, which records complete marriage histories.}

Table \ref{tab-prewar-diff} reports means and standard deviations of prewar covariates for war widows (Columns (1) and (2)) and non-widows (Columns (3) and (4)). %
Column (5) reports t-statistics for testing the null of no differences in covariate means between the two groups. Given the large sample size of 30,351, even small differences will be statistically significant. Therefore, Column (6) also reports normalized differences, which scale differences in means by the square root of the sum of the variances to assess overlap in the covariate distribution \citep{Imbens2009}.

All normalized differences in prewar covariates %
are smaller than 0.10, except for two: age and displacement status. War widows tend to be slightly older than non-widows; as earlier cohorts were more likely to be married before WWII, they were  more likely to lose their husband. War widows were also somewhat more likely to be displaced than non-widows, presumably because of the elevated death rates of soldiers from Germany's former eastern territories \citep{Overmans1999}. However, even these differences are small so that they can be (robustly) controlled for in standard linear regressions.%

\begin{table}[h!]
	\begin{threeparttable}
		\caption{Prewar differences between later war widows and non-widows} \centering
		\label{tab-prewar-diff}
		\begin{footnotesize}
			\begin{tabular}{lcccccccc}
				\toprule\toprule
				& \multicolumn{2}{c}{War widows} && \multicolumn{2}{c}{Non-widows} &&  &  \\
				& \multicolumn{2}{c}{(N=5,129)} && \multicolumn{2}{c}{(N=25,222)} &&  \multicolumn{2}{c}{Differences}  \\  \cline{2-3} \cline{5-6} \cline{8-9}
				& mean	& st.d.	&& mean	& st.d. && t-stat & norm. \\		
				& (1)	& (2)	&& (3)  & (4) && (5) & (6)  \\
				\midrule
			\multicolumn{9}{l}{\textit{Socio-demographic characteristics}:} \\ 
			Birth year 					& 1910.331 	& 2.479 && 1909.889 & 2.556	&& 11.349 	& 0.176 \\
			House ownership (0/1)		& 0.488		& 0.500	&& 0.466	& 0.499	&& 2.877	& 0.044 \\
			Years of education			& 8.944		& 1.580	&& 9.046	& 1.655	&& 4.039	& 0.063 \\	
			Years of schooling			& 8.356		& 1.065 && 8.408	& 1.148	&& 3.044	& 0.048 \\
			Siblings					& 4.807		& 2.731	&& 4.674	& 2.724 && 3.182	& 0.049 \\ \\
			
			\multicolumn{9}{l}{\textit{Place of residence}:} \\
			Eastern Europe (incl. eastern territories) & 0.208 & 0.406 && 0.155 & 0.362 && 9.293 & 0.137 \\
			Soviet occupation zone					   & 0.047 & 0.212 && 0.052 & 0.221 && 1.319 & 0.020 \\ \\
			
			\multicolumn{9}{l}{\textit{Employment and occupational status (\%):}} \\
			 Employed 					& 0.419		& 0.494	&& 0.388	& 0.487 && 4.233	& 0.065 \\
			 \; Market employment 		& 0.331		& 0.471	&& 0.289	& 0.453	&& 6.064	& 0.092 \\
			 \;\;\; Self employed$^1$	& 0.025	    & 0.157 && 0.022	& 0.146	&& 1.421	& 0.021 \\
			 \;\;\; Farmer$^2$			& 0.023		& 0.151 && 0.012	& 0.111 && 5.894 	& 0.081 \\
			 \;\;\; Civil servant		& 0.003		& 0.056 && 0.003	& 0.058 && 0.240	& 0.004 \\
			 \;\;\; White collar		& 0.082		& 0.274	&& 0.091	& 0.288 && 2.119	& 0.033 \\
			 \;\;\; Blue collar			& 0.197		& 0.398	&& 0.159	& 0.366	&& 6.651	& 0.099 \\
			 \;\;\; Apprentices			& 0.001		& 0.031	&& 0.001	& 0.030 && 0.135	& 0.002 \\
			 \; Helping family			& 0.088		& 0.284	&& 0.099	& 0.298	&& 2.364	& 0.037 \\
			 In education				& 0.002 	& 0.044 && 0.003	& 0.052	&& 1.007	& 0.016 \\
			 Unemployed					& 0.001		& 0.024	&& 0.001	& 0.026	&& 0.227	& 0.004 \\
			 Out of the labor force		& 0.578		& 0.494	&& 0.609	& 0.488	&& 4.110	& 0.063 \\ \\
			 
			 \multicolumn{9}{l}{\textit{Sector of employment (\%):}} \\
			 Agriculture				& 0.121		& 0.326 && 0.099	& 0.299	&& 4.707	& 0.070 \\
			 Industry					& 0.143		& 0.350	&& 0.123	& 0.329	&& 3.918	& 0.059 \\
			 Construction				& 0.002		& 0.039	&& 0.003	& 0.054	&& 1.689	& 0.028 \\
			 Trade						& 0.073		& 0.261	&& 0.079	& 0.270	&& 1.427	& 0.022 \\
			 Finance					& 0.013		& 0.114	&& 0.016	& 0.127	&& 1.714	& 0.027 \\
			 Services					& 0.066		& 0.249	&& 0.065	& 0.247	&& 0.302	& 0.005 \\
			 Not employed or unknown	& 0.581		& 0.493	&& 0.614	& 0.487	&& 4.353	& 0.066 \\
			 \bottomrule\bottomrule   
			\end{tabular}%
		\end{footnotesize}
		\begin{tablenotes}
			\item \footnotesize{\emph{Notes}: Sample means and standard deviations of prewar covariates for war widows and non-widows in our baseline sample (after dropping observations with missing prewar characteristics). All data refer to 1939 except for education, schooling, and the number of siblings, which are measured in 1971. $^1$ Self-employed outside agriculture. $^2$ Farmer with own land. The t-statistic in Column (5) refers to a two-sided mean difference t-test. Normalized differences in Column (6) are calculated as $|\bar{X}_{1}-\bar{X}_{0}|/\sqrt{(S_1)^2+(S_0)^2}$ where $\bar{X}_{1}$ and $\bar{X}_{0}$ are the sample means and $(S_1)^2$ and $(S_0)^2$ the sample variances of war widows and non-widows, respectively.}
		\end{tablenotes}
	\end{threeparttable}
\end{table}

\paragraph{German Life History Study (GHS).} 

For more detailed life-cycle analysis, we use the GHS, a retrospective survey of eight West German birth cohorts. We draw on the second wave (GHS-2, conducted in 1985-88), which surveyed 1,412 respondents born in 1919-21, of whom 853 are women \citep{ZA2647,ZA2646}. While the sample size is much smaller, the GHS %
contains the respondents' complete marriage, education, employment, and occupational history.\footnote{We use the Standard International Occupational Scale (Treiman, 1977) to study occupational success, with scores ranging from 18 (unskilled laborers) to 78 (medical professionals, professors).} %
We can thus study war widows' labor market outcomes over the entire life cycle. %
We classify women whose husbands died in or before 1945 ($N=95$) as war widows. The control group are women %
who also married in or before 1945 but did not lose their husbands during the war ($N=428$). 

\paragraph{ALLBUS.} For complementary evidence on work attitudes and gender norms, we consider ALLBUS, a survey of the attitudes, behavior, and social structure of the German population that has been conducted biannually since 1980. The data allow us to examine the work attitudes of war widows and their children. Appendix \ref{sec:norms} describes the data in detail.

\subsection{Empirical Strategy}
\label{Section Empirical Strategy}
To examine whether widows and non-widows who were comparable before the war fared differently after 1945, we estimate OLS regression models of the following type:
\begin{equation}
	y_{it}=\alpha+\bm{x}_{i,39}\bm{\beta} +\delta D_{i}+\epsilon_{it},\label{eq:est1}
\end{equation}
where $y_{it}$ is a particular postwar outcome of person $i$ at time $t$, $D_{i}$ is a dummy variable identifying war widows, $\bm{x}_{i,39}$ is a row vector of prewar control variables, %
and $\epsilon_{it}$ is an error term. Our main parameter, $\delta$, measures the ``widowhood effect,'' i.e., the average difference in a given outcome between war widows and otherwise (as of 1939) comparable non-widows.%
\footnote{The large number of widows may have affected the economy as a whole and thus also the control group. Our regression does not capture such general equilibrium effects. However, from a policy perspective, the relative economic fortunes of war widows are of primary interest, not the situation that would have prevailed in their absence.} We report robust standard errors, clustered at selection districts.\footnote{The survey drew a sample of districts, interviewing all persons of the selected districts \citep{Tegtmeyer1979}.} %
We also compare the OLS estimates to estimates based on inverse probability weighting (IPW).\footnote{IPW models the treatment rather than the outcome \citep{Imbens2009}. It estimates the widowhood effect by comparing \textit{weighted} outcome means of widows and non-widows, giving higher weights to control group observations with a high probability of losing their husbands based on prewar covariates.}

Identification requires that conditional on $\bm{x}_{i,39}$, widowhood status $D_{i}$ is uncorrelated with unobserved prewar differences that still affect economic outcomes in postwar Germany. This assumption would be violated if women with better unobserved labor market skills married high-skilled men who were less likely to die in WWII. Non-widows then had better labor market prospects on average, which would explain why we do not find any lasting impact of widowhood on employment. Several pieces of evidence speak against this concern. 

First, women in our sample were predominantly married to men whose cohorts were largely or entirely conscripted for the war. This is important because differences in conscription rates, rather than unequal survival rates, are the main explanation for why mortality rates differed between birth cohorts \citep{Overmans1999}.\footnote{While skilled workers in the armaments industry were initially spared military service \citep{Mueller2016}, more than 80 percent of German military deaths occurred after 1942 \citep{Overmans1999}, when all reserves were mobilized.} Importantly, our results also apply to younger women born 1915-21 (see Appendix \ref{sec:altcontrol}), almost all of whom were married to men born 1910-25. Their cohorts were fully conscripted during the war, so that selection into military service played a negligible role. %
These cohorts were also far too young for middle and higher officer ranks, which might have promised better survival prospects. %
Consistent with these arguments, \cite{Braun2023} show that socioeconomic background does not predict war service or injuries among war survivors born in 1919-21.%

Second, Table \ref{tab-prewar-diff} documents only small prewar differences between widows and non-widows for our main MZU71 sample (explaining merely 1\% of the variation in widowhood status, see Appendix Table \ref{tab:predwarwidow}). %
Third, Appendix Table \ref{tab:exoshocks} shows for the GHS auxiliary sample that war widowhood is not only uncorrelated with women's own prewar characteristics,\footnote{Widowhood does correlate with the marriage year, as women who marry earlier face a longer risk of losing their spouse. %
Therefore, we control for the marriage year in all GHS regressions. While we cannot control for the marriage year in the MZU71, we restrict this sample to women who were well above the average age of first marriage in 1945.} but also with their spouse's and parents' characteristics. Fourth, we show below that adding the extensive set of prewar covariates listed in Table \ref{tab-prewar-diff} has virtually no effect on our estimates, while helping to explain variation in the outcome variables. It is thus unlikely that omitted variables drive our results.%

\section{Socioeconomic Effects of WWII Widowhood}
\label{sec:results}

\subsection{Sociodemographic Outcomes}

Table \ref{tab-income-wealth} shows the effect of war widowhood on sociodemographic outcomes in the MZU71, a quarter century after the war's end. The women in our sample were 56-65 years old at that time. Column (2) presents estimates from a parsimonious OLS model, controlling only for age. Column (3) adds control variables for house ownership in 1939, the displacement status,  %
the number of siblings, and a full set of education dummies. The full-fledged specification in Column (4) additionally accounts for the sectoral and occupation affiliation in 1939. Finally, Column (5) presents IPW estimates using the full set of controls to predict treatment status. In what follows, we discuss the point estimates from the IPW model, but the estimates hardly change across specifications. This supports the unconfoundness assumption we invoke for identification.

\begin{table}[h!]
	\begin{center}
		\begin{threeparttable}
			\caption{The impact of WWII widowhood on demography, labor market status, income, and wealth in 1971} \centering
			\label{tab-income-wealth}
			\begin{small}
				\begin{tabular}{lcccccc}
					\toprule\toprule
					&Control  & &&& &Obser-\\
					&mean	 & \multicolumn{3}{c}{OLS} & IPW &  vations\\ \cline{3-5}
					&  (1)	&(2)	& (3)	& (4)	& (5) & (6) \\ \midrule
					\multicolumn{7}{l}{\textit{A. Demographic outcomes}:} \\				
					Married (0/1)			& 0.664	& -0.488	& -0.489	& -0.485	& -0.486 & 30,251\\
					&		&(0.007)	& (0.007)	& (0.007)	& (0.007) &\\
					Living with non-relatives (0/1) & 0.010 & 0.012 & 0.012 & 0.012 &0.012 & 30,351 \\
					&& (0.002) & (0.002) & (0.002) & (0.002) &  \\
					Living alone (0/1)		& 0.246 & 0.335 	& 0.338		& 0.335		& 0.335 & 30,351 \\
					&		& (0.008)	& (0.008)	& (0.008)	& (0.008) &\\ 	
					Number of children		& 2.130	& -0.253	& -0.287	& -0.278	& -0.282 & 30,351 \\	
					&		& (0.025)	& (0.025)	& (0.024)	& (0.024) & \\ 		
					\multicolumn{7}{l}{\textit{B. Labor market status}:} \\
					Market employment	 & 0.162  	& -0.018 	&  -0.014	& -0.018 	&  -0.019 & 30,351 \\
					& 			& (0.006)	& (0.006)	&  (0.006)    & (0.006) & \\			
					Helping family member & 0.061	& -0.033	& -0.033	& -0.031	& -0.031 & 30,351 \\
					&			& (0.003)	& (0.003)	& (0.003)	& (0.003) & \\		
					Out of the labor force &  0.776	& 0.052 	& 0.048 	& 0.050 	&  0.051 & 30,351 \\
					&	& (0.006)	&  (0.006)	&   (0.006)    & (0.006) &\\					 			
					\multicolumn{7}{l}{\textit{C. Income}:} \\
					Personal income			& 241.018 & 243.295 & 250.198 & 248.311 & 246.919 & 28,763 \\
					\footnotesize{(unconditional)}				&&(5.768)	& (5.459)	& (5.435) & (5.423) & \\
					Personal income & 676.935 & 118.043 & 98.186 & 104.075 & 105.252 & 4,561\\
					\footnotesize{(conditional on market work)} && (18.325) & (16.083) & (15.690) & (15.882) &\\ 
					Welfare support as main income	& 0.297 & 0.454 	& 0.451 	& 0.448 	& 0.449 & 30,351 \\
					& 		& (0.007) 	& (0.007) 	& (0.007) 	& (0.007) & \\ 
					Household income, square-root scale$^a$	& 765.377 & -64.782	& -50.846	& -45.984	& -46.630 & 26,673 \\
					& 	&(6.881)	& (6.587)	& (6.515)	& (6.485) & \\ 
					\multicolumn{7}{l}{\textit{D. Wealth}:} \\
					House ownership			& 0.406	& -0.096 	& -0.095	& -0.095	& -0.095 & 30,351 \\
					&		&(0.008)	& (0.007)	& (0.007)	& (0.007) & \\ \bottomrule
					Sociodemographic controls & &no & yes & yes & yes & \\
					Labor market controls	  & &no & no & yes & yes & \\				
					\bottomrule\bottomrule   
				\end{tabular}%
			\end{small}
			\begin{tablenotes}
				\item \footnotesize{\emph{Notes}: Means of the control group and estimates for war widowhood. Each estimate stems from a separate regression. Estimates in Columns (2)-(4) are by OLS, estimates in Column (5) by IPW. Regressions include the following prewar covariates: (2) full set of age dummies, (2) = (1) plus an indicator for house ownership in 1939, indicators for expellees from Eastern Europe and refugees from GDR, number of siblings, full set of education dummies, (3)/(4) = (2) plus seven categories for the sector of employment in 1939 (agriculture, industry, construction, trade/transport, finance, public and private services, unknown) and seven categories for the occupational or employment status in 1939 (self-employed, farmer, civil servant, white-collar worker, blue-collar worker, helping family member, out of the labor force including apprentices, in education, and unemployed). Robust standard errors clustered at selection districts (\textit{Auswahlbezirke}) are reported in parentheses. $^a$ The square-root scale divides total household income by the square root of household size. We do not observe younger household members born 1956 or later in the MZU71.}
			\end{tablenotes}
		\end{threeparttable}
	\end{center}	
\end{table} 

Panel A. of Table \ref{tab-income-wealth} shows that war widowhood had striking long-term demographic consequences. By 1971, war widows have a 48.6 pp lower probability of being married (from a baseline probability of 66.4\%, see Column (1)). Finding a new partner often proved elusive, as the war led to an acute shortage of men.  Although war widows were more likely to cohabit with a non-family member, the overall probability was low, about 2\%. Thus, we find little evidence to support the widespread belief that many war widows cohabited without marrying in order to retain their pension rights.\footnote{All adult household members are legally obligated to provide information for the microcensus, ensuring that non-response is not an issue.} War widows were also more than twice as likely to live alone, but had only slightly fewer children (as women in our sample often had children before 1945).

\begin{figure}[!tbp]
	\caption{Life-cycle effects of war widowhood (GHS)}\label{fig:GHS}
	\centering
	\begin{threeparttable}
		\begin{center}
			\textit{\begin{center}
					Socio-demographic characteristics\\ \medskip
			\end{center}}
			\subfloat[Married]{\includegraphics[clip, trim=0.3cm 0.7cm 0.3cm 0.7cm, width=0.47\textwidth]{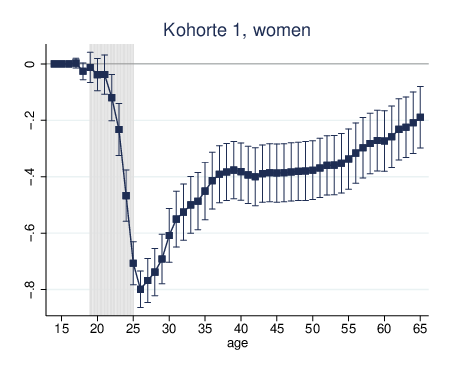}\label{fig:GHS_married}}
				\hfill
			\subfloat[\# of children]{\includegraphics[clip, trim=0.3cm 0.7cm 0.3cm 0.7cm, width=0.47\textwidth]{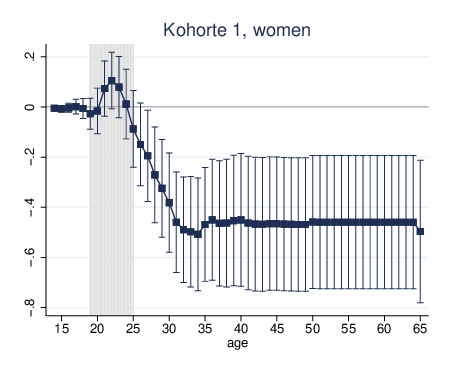}\label{fig:GHS_kids}}\\
			\textit{\begin{center}
					Employment and career outcomes\\ \medskip
			\end{center}}
			\subfloat[Employment]{\includegraphics[clip, trim=0.3cm 0.7cm 0.3cm 0.7cm, width=0.47\textwidth]{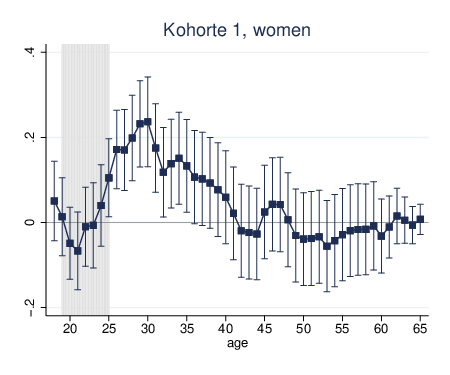}\label{fig:GHS_emp}}
			\hfill
			\subfloat[Occupational score (when employed)]{\includegraphics[clip, trim=0.3cm 0.7cm 0.3cm 0.7cm, width=0.47\textwidth]{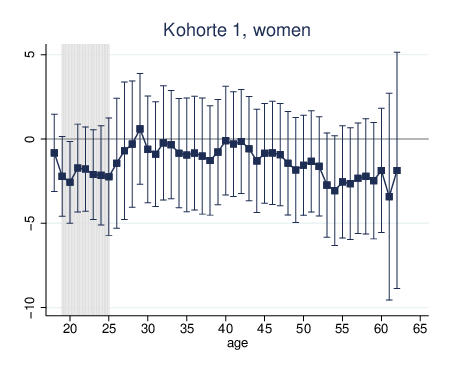}\label{fig:GHS_occ}}\\
			\textit{\begin{center}
					Intergenerational spillovers\\ \medskip
			\end{center}}
			\subfloat[Sons of war widows]{\includegraphics[clip, trim=0.3cm 0.7cm 0.3cm 0.7cm, width=0.47\textwidth]{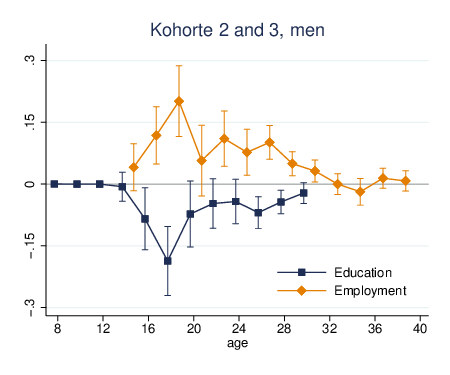}\label{fig:GHS_igmen}}
			\hfill
			\subfloat[Daughters of war widows]{\includegraphics[clip, trim=0.3cm 0.7cm 0.3cm 0.7cm, width=0.47\textwidth]{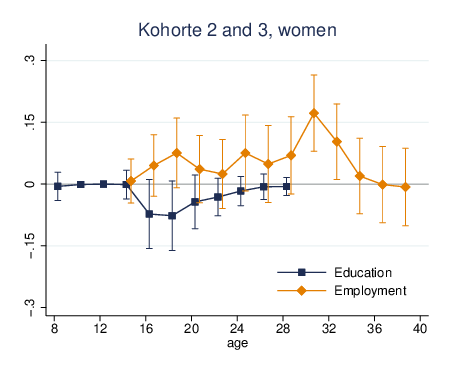}\label{fig:GHS_igwomen}}\\
						\textit{\begin{center}
			\end{center}}			
		\end{center}
		\vspace*{-5mm}
		\begin{tablenotes}
			\item \footnotesize{\emph{Notes}: GHS, estimated effect of war widowhood over the life cycle for women born 1919-21. Estimates are from a pooled OLS regression, interacting indicators for war widows (panels (a)-(d)) or their children (panels (e)-(f)) and birth year with a full set of age indicators. Panels (a)-(d) also control for the year of marriage. Point estimates are marked by a dot. The vertical bands indicate the 95\% confidence interval of each estimate. The shaded area indicates the duration of WWII.}
		\end{tablenotes}
	\end{threeparttable}
\end{figure}

The top panels of Figure \ref{fig:GHS} illustrate these demographic consequences over the life cycle, based on the GHS and a pooled OLS regression in which we interact a full set of age indicators with the indicator for war widows. %
Figure \ref{fig:GHS}(a) shows that war widows %
were 80 pp less likely to be married by the war's end. This gap then shrinks as some widows remarry, %
but stabilizes at around 40 pp when widows reach their late 30s. The gap starts shrinking again %
when widowhood becomes more common also in the control group. %
Figure \ref{fig:GHS}(b) illustrates that war widows had fewer children only after the war ended. %

\subsection{Labor Market Outcomes}

Panel B. of Table \ref{tab-income-wealth} shows that in the MZU71, war widows were \textit{less} likely to be employed in 1971, having a 1.9 pp lower probability to be in market employment, a 3.1 pp lower probability to be a helping family member, and thus a 5.1 pp higher probability to be out of the labor force than non-widows. %
This finding is surprising, as most war widows remained unmarried, and unmarried women had considerably higher participation rates at the time. %

To better understand these patterns, Table \ref{tab-lab-outcomes} reports the effect of war widowhood on employment and occupational status in 1950, 1960, and 1971. %
We focus on IPW estimates %
with the full set of controls. %
Panel A. shows that %
in 1950, war widows were 13.8 pp more likely to perform market work than non-widows with similar prewar characteristics (from a baseline probability of 20.5\%). However, the difference shrinks to 4.7 pp by 1960, and by 1971 war widows were less likely to work. %
These life-cycle patterns cannot be explained by changes in aggregate labor demand, which would affect both treatment and control groups. The negative long-term effect on labor force participation is reinforced by the fact that, in all years, war widows were only half as likely as non-widows to work as helping family members. %

Appendix \ref{subsec:het} highlights that women who were not in market work before the war were particularly drawn into the labor force in 1950. Among those who were helping family members, war widows were 26.1 pp (or nearly 300\%) more likely to be in market employment in 1950. But even for this group, the positive employment effect eventually disappears. Moreover, highly educated women were much more likely to enter employment after their spouse's death than low-educated women, with a modest positive employment effect persisting into old age. This aligns with US findings that WWII mobilization increased employment only among highly educated women \citep{Goldin2013}. The negative employment effect at old age is driven by women with children (see again Appendix \ref{subsec:het}). 

Figure \ref{fig:GHS}(c) confirms these life-cycle patterns in the GHS, showing higher employment among war widows compared to non-widows in the postwar period, but similar or lower employment rates by their mid-40s and into retirement. Widows reduce their labor supply at the same age as the share of remarried widows increases (Figure \ref{fig:GHS}(a)). Appendix Figure \ref{fig:remarriage_eventstudy} confirms that this is no coincidence: widow's employment rate drops by 20 pp immediately after remarriage. However, in the long run, we also observe a drop in employment for war widows in the GHS who did not remarry and for the older cohorts of widows in the MZU71 who rarely remarried (see Table \ref{tab-income-wealth}). %
Despite their higher employment rates at early age, war widows reported similar average occupational prestige scores than non-widows (see Figure \ref{fig:GHS}(d)). %

\begin{table}[h!]
	\begin{center}
	\begin{threeparttable}
		\caption{The impact of war widowhood on labor market outcomes in 1950-1971} \centering
		\label{tab-lab-outcomes}
		\begin{small}
			\begin{tabular}{lcccccccc}
				\toprule\toprule
				&\multicolumn{2}{c}{1950} &&\multicolumn{2}{c}{1960}  &&\multicolumn{2}{c}{1971} \\ \cline{2-3} \cline{5-6} \cline{8-9}
				&Control  & 	&& Control	&		&& Control	& \\
				&mean	  & IPW && Mean		& IPW	&& mean		& IPW \\ 
				&  (1)	&(2)	&& (3)	& (4)	&& (5) & (6)\\ \midrule
				\multicolumn{6}{l}{\textit{A. Labor market status}:} \\				
				Market employment 		& 0.205	& 0.138		&& 0.238 & 0.047	&& 0.162	& -0.019 \\
								  	& 		& (0.007)	&& 		 & (0.007)	&& 			& (0.006) \\
				Helping family			& 0.089	& -0.045	&& 0.084 & -0.044	&& 0.061	& -0.031 \\
									&		& (0.003)	&&		 & (0.003)	&&			& (0.003) \\
				Unemployed			& 0.004	& 0.000		&& 0.000 & 0.000	&& 0.001	& -0.001 \\		 	
									&		& (0.001)	&&		 & (0.001)  && 			& (0.000) \\
				Out of the labor force &0.703 & -0.093	&& 0.677 & -0.003	&& 0.776	& 0.051 \\
									&		& (0.007)	&& 		 & (0.007)	&&			& (0.006) \\ \hline 
				Observations		& 		&29,302	&	&& 	30,342	   &&&	30,351		  \\ \\

				\multicolumn{6}{l}{\textit{B. Occupational status (conditional on market work)}:} \\				
				Self employed			& 0.135	& -0.023	&& 0.128 & -0.013	&& 0.132	& -0.012 \\
									& 		& (0.007)	&& 		 & (0.008)	&& 			& (0.012) \\
				Farmer				& 0.055	& 0.017		&& 0.045 & 0.017	&& 0.039	& 0.014 \\
									&		& (0.006)	&&		 & (0.006)	&&			& (0.008) \\
				Civil servant			& 0.009	& 0.005		&& 0.010 & 0.010	&& 0.010	& 0.018 \\		 	
									&		& (0.002)	&&		 & (0.003)  && 			& (0.005) \\
				White collar 			& 0.249 & 0.012		&& 0.265 & 0.022	&& 0.302	& 0.031 \\
									&		& (0.010)	&& 		 & (0.011)	&&			& (0.016) \\
				Blue collar				& 0.552 & -0.013	&& 0.551 & -0.036	&& 0.517	& -0.050 \\
									& 		& (0.011)	&&		 & (0.011)	&&			& (0.016) \\ \hline
				Observations		& 		&6,783	&	&& 		7,547   &&&			4,880  \\

				\bottomrule\bottomrule   
			\end{tabular}%
		\end{small}
		\begin{tablenotes}
			\item \footnotesize{\emph{Notes}: Means of the control group and IPW estimates for war widowhood. Each estimate stems from a separate regression. Regression include as controls a full set of age dummies, an indicator for house ownership in 1939, indicators for expellees from Eastern Europe and refugees from GDR, the number of siblings, a full set of education dummies, indicators for the sector of employment in 1939 (agriculture, industry, construction, trade/transport, finance, public and private services, unknown) and the occupational or employment status in 1939 (self-employed, farmer, civil servant, white-collar worker, blue-collar worker, helping family member, out of the labor force including apprentices, in education, and unemployed). Robust standard errors clustered at selection districts (\textit{Auswahlbezirke}) are reported in parentheses.}
		\end{tablenotes}
	\end{threeparttable}
	\end{center}	
\end{table}

Panel B. of Table \ref{tab-lab-outcomes} shows evidence on the occupational status from our larger main sample. Conditional on market work, war widows were less likely to work as blue-collar workers. %
In 1971, war widows were 5.0 pp less likely to be blue-collar workers (baseline probability: 51.7\%). Instead, war widows were overrepresented in white-collar occupations. %
They were also more than 2.5 times as likely to be employed as civil servants (albeit from a baseline of only 1.0\%), likely as a result of policies that favored war widows %
for civil service employment. %

Panel C. of Table \ref{tab-income-wealth} shows that the personal net income of war widows was about twice that of non-widows, largely due to higher support by the state: war widows were 2.5 times more likely to report welfare support as their main income source (baseline probability: 29.7\%). If we focus on women in market work, war widows have only DM 105 more than non-widows (baseline: DM 677). This is considerably less than the unconditional basic pension of DM 198 paid to war widows at the time. Panel C. also shows that household income, measured at the square root equivalence scale, is about 6\% lower for war widows than for non-widows.%

Finally, Panel D. shows that %
war widowhood reduced the probability of owning a house in 1971 by 9.5 pp (or 23.4\%).%

\subsection{Mechanisms}\label{sec:mechanism}

Despite an initial increase in employment, war widows eventually had lower employment rates than non-widows. In their 20s and 30s, they thus bore the double burden of work and childcare without a partner. However, in their 40s and 50s, after their children had left home, war widows were less likely to work than non-widows. 
We discuss here the mechanisms %
that might have contributed to this counterintuitive life-cycle pattern.

To fix ideas, we consider a simple static model of labor supply in which mothers choose consumption $c$, hours of work $h$, leisure time $l$ and child-care time $t$ to maximize the utility function 
\[
\underset{l,t}{max}~U(c(l,t),l,q(t);\theta),
\]
subject to a ``disutility of work'' $\theta$, which might vary with age or social norms, %
and constraints for time $l_{0}=h+l+t$, %
child ``quality'' $q=q(t)$ with $q'>0$ and $q''<0$, and the budget, 
\[
c=w\underset{=h}{(\underbrace{l_{0}-l-t})}+R_{0},
\]
where $w$ is the hourly wage rate and $R_{0}$ are other sources of household income. %
The optimal choices of leisure and work time are characterized by the slopes of the indifference curves between consumption and leisure and between consumption and child ``quality'', and the budget constraint (see Appendix \ref{sec:appendix_model} for details). %

Although simple, the model can rationalize the peculiar life-cycle labor supply trajectory we observe. Similar to \cite{Boehnke2022}, we interpret the loss of a spouse as a negative income shock, i.e., a decrease in $R_{0}$. Assuming that leisure is a normal good, such income loss decreases leisure and time spent on childcare, and increases hours worked (\emph{income effect}). Appendix Figure \ref{fig:model}a provides an illustration. 

War widows also received compensation %
(see Section \ref{Section Background}). But, since this offsets only part of the decline in $R_{0}$ (household income decreases, as shown in Table \ref{tab-income-wealth}), this would not yet explain why war widows were \emph{less} likely to work at older ages. However, as the widow pensions were partially means-tested, they also decrease the effective take-home wage, which further reduces work incentives (\emph{substitution effect}). 

Crucially, the means-tested form of compensation grew disproportionately over time, accounting for 72\% of maximum pensions in 1970. %
Therefore, the budget curve of war widows with respect to hours worked became flatter over time, disincentivizing their participation in the labor force. The implied (negative) substitution effect on hours worked can dominate the (positive) income effect from a reduction in total income, as illustrated in Appendix Figure \ref{fig:model}b. 

The compensation scheme thus created ``perverse'' life-cycle incentives for war widows. Low compensation in earlier years forced them into the labor force, despite limited childcare and the stigma that society placed on working single mothers. Once their children reached adulthood, war widows had more time available, but the increasingly means-tested compensation scheme discouraged work. A less aggressively means-tested compensation scheme could have maintained labor force participation, reduced public spending, and provided similar utility to war widows (see Appendix Figure \ref{fig:model}b).

The negative work experience of the 1950s, with the double burden of work and childcare in an environment hostile to working mothers, may also explain why war widows were not more supportive of women with young children working than non-widows, although they tended to have more progressive gender norms in other domains. As we show in Appendix \ref{sec:norms}, about nine in ten war widows, the same as among non-widows, disapproved of women working outside the home when there are small children at home. War widows also did not place more importance on work in their lives. This probably contributed to their withdrawal from the labor market, as did the physical exhaustion from the double burden of work and childcare at a younger age.\footnote{Consistent with this interpretation, war widows report higher levels of illness than the control group, and this gap increases with age; see Appendix Figure \ref{fig:illness}.}

\section{Intergenerational Spillovers}
We can also examine the intergenerational spillovers of war widowhood, as the first wave of the GHS (GHS-1), conducted in the early 1980s, interviewed children born shortly before (1929-31) and during (1939-41) the war \citep{ZA2645}. The treatment group consists of those whose father (but not mother) died during WWII. The control group includes children from the same cohorts whose fathers were absent for at least one year during the war. For educational outcomes, we can also include the respondent's siblings, as the GHS-1 provides educational information for close family members.
\textit{}
We use this information to estimate a variant of equation (\ref{eq:est1}), 
\begin{equation}
	y_{it}=\alpha+\bm{x}_{i,39}\bm{\beta} +\delta_{m} D_{i}+\delta_{gap} D_{i} 1(i=\text{female}) +\epsilon_{it},\label{eq:est2}
\end{equation}
where $D_{i}$ is a dummy variable indicating whether the father of individual $i$ died in the war, $\delta_{m}$ measures the intergenerational spillover on the male children of war widows, and $\delta_{gap}$ the differential impact on daughters relative to sons. %
The controls $\bm{x}_{i,39}$ include the mother's birth year and interactions between indicators of the child's birth year and gender. 

Table \ref{tab-intergenerational} reports our estimates. Column (1) shows that the sons of war widows had half a year less schooling, in line with previous findings in \cite{Ichino2004}, who report negative effects of fathers' WWII service and death on children's years of schooling. Tertiary schooling (university or vocational training) is even 30\% lower in the treatment group (Column (2)). Overall, educational attainment declines by one year (Column (3)). In contrast, there is little decline in the educational attainment among the daughters of war widows; the estimated effect on the gender gap $\delta_{gap}$ is of the opposite sign and almost as large as the main effect $\delta_{m}$.%

\begin{table}[h!]
	\begin{center}
		\begin{threeparttable}
			\caption{Intergenerational spillovers on the children of war widows} \centering
			\label{tab-intergenerational}
			\begin{small}
				\begin{tabular}{lcccccccc}
					\toprule\toprule
					& \, Schooling \,	& Tertiary 		& Schooling      & \multicolumn{2}{c}{Employment (years)} && \multicolumn{2}{c}{Occupational scores}  \\ \cline{5-6} \cline{8-9} %
					& 		       	&  schooling  & w/ tertiary 	& age 16-29 	& age 30-39 & & own & spouse   \\%	& prestige  \\
					&  (1)		& (2) 		& (3)			& (4)			& (5)	&	&  (6)        &    (7)    \\ \midrule%
					Father's death&      -0.478&      -0.524&      -1.005&       1.337&       0.080  &&  -0.704   &   0.736  \\
            				&     (0.127)   &     (0.156)   &     (0.217)   &     (0.399)   &     (0.118)  &  &   (1.643)  &  (2.454)   \\
					Father's death&       0.347  &       0.476 &       0.828&      -0.451   &       0.260 &&   -0.972   &   -2.227   \\
 					\, x daughter           &     (0.187)   &     (0.193)   &     (0.296)   &     (0.665)   &     (0.594) &&  (2.246)  & (3.115)  \\
					Control mean        &       8.744   &       1.722   &      10.468   &       8.597   &       6.839  &&  41.973  &  39.708 \\
					Observations           &    1,722   &    1,718   &    1,718   &     615   &     615    &&   615   &   563    \\			
					\bottomrule\bottomrule   
				\end{tabular}%
			\end{small}
			\begin{tablenotes}
				\item \footnotesize{\emph{Notes}: Estimates of the effect of war widowhood on educational attainment, employment (in years) or occupational scores (maximum score) of their children and partners. Each estimate stems from a separate OLS regression including the mother's birth year and interactions between indicators of the child's birth year and gender as control variables. GHS birth cohorts 1929-31 and 1939-41 (including siblings in Columns (1)-(3)). Robust standard errors clustered at the household level are reported in parentheses.}
			\end{tablenotes}
		\end{threeparttable}
	\end{center}
\end{table} 

With less time spent in education, the children of war widows instead enter the labor force earlier, presumably because of the financial hardship they faced after the war. As shown in Column (4), the sons of war widows spent nearly 1.3 years more in employment during age 16-29 than the control group. %
Figure \ref{fig:GHS}(e) and (f) confirm that the decline in educational attainment is mirrored by a corresponding increase in employment. Despite the lower educational attainment of male children, we find no spillovers on their employment at mid-age (Table \ref{tab-intergenerational}, Column (5)) or their occupational scores (Column (6)).

Importantly, we find no persistent employment effect for the daughters of war widows either and, if anything, a negative effect on their occupational success. %
While they are more likely to work in their 20s, this gap disappears by their mid-30s. We also find no positive spillovers on work attitudes: if anything, the daughters of war widows held \textit{less} progressive views on the compatibility of women's work and childcare responsibilities (see Appendix \ref{sec:norms}). We also find no spillovers to their sons' gender norms (Appendix \ref{sec:norms}) or to the occupational scores of their sons' \textit{wives} (Table \ref{tab-intergenerational}, Column (7)). While \cite{Fernandez2004} show that women's labor force participation can affect their sons' preference for a working wife, we find no such spillovers for the sons of war widows. 

Our results complement previous work on the long-term effects of the two world wars on women's employment. While war-induced increases in women's employment can transform work and gender norms and trigger intergenerational spillovers \citep{Fernandez2004,Fernandez2013,Gay2023}, we find no such effects in our treatment group: War widows did not have more positive attitudes toward women's work, and did not transmit more progressive attitudes to their sons or daughters--likely reflecting the challenging conditions they endured when working as single mothers in the postwar period.\footnote{\cite{Bastiaans2023} propose a similar interpretation to explain why in the Netherlands, a policy-induced increase in labor supply among women with low labor force attachment decreased their daughter's labor supply.} Employment spells born of necessity (such as for war widows) thus appear to have less lasting effects than those resulting from higher aggregate labor demand (due to men's wartime mobilization).

\section{Conclusion}\label{sec:conclusion}

Millions of women worldwide have lost their husbands in violent conflicts, yet little is known about their economic situation. This includes WWII, the most devastating conflict in history. We present first evidence on the complex effects of war widowhood, using life course data from postwar Germany. War widowhood had dramatic long-term consequences for family formation, with most widows remaining unmarried after 1945. Despite the persistent negative shock to household income, war widows were more likely to work only into their middle years. Over the life cycle, the positive employment effect gradually diminished, and by 1971 war widows were less likely to work and 2.5 times more likely to rely on welfare than their peers.

Our findings underscore the importance of public policy in shaping widows' economic outcomes. Low compensation and inadequate child care in the immediate postwar period created a dilemma for war widows with young children who needed earned income but faced accusations of neglecting their children if they worked. The dire financial situation likely also forced widows' children to leave school early. Compensation then became more generous and increasingly means-tested over time, creating disincentives to work long after the war ended. The poor work experiences of war widows in the 1950s may also explain why their war-induced labor market entry did not have spillover effects on the participation of the next generation.

Given the often dire financial straits of governments in conflict zones, the problem of insufficient support for war widows is not specific to our study, but applies similarly to contemporary conflicts \citep[see e.g.][on South Sudan's civil war and the Russo-Ukrainian war, respectively]{Mednick2018,York2023}. International efforts to address the challenges faced by war widows are thus particularly important during and immediately after conflicts, when many widows face the double burden of childcare and employment. %
Such support is also critical to break the intergenerational reproduction of conflict-related poverty in widow-headed households \citep{Buvinic2013}, which our study suggests may operate through reduced educational opportunities. Finally, as countries' fiscal space expands with postwar recovery, it is important to design widow compensation in a way that preserves incentives to work. %

	\singlespacing
	
	\bibliographystyle{apalike}
	\begin{small}
		\bibliography{literature-widows}

\begin{thebibliography}{}

\bibitem[Abramitzky et~al., 2011]{Abramitzky2011}
Abramitzky, R., Delavande, A., and Vasconcelos, L. (2011).
\newblock Marrying up: {T}he role of sex ratio in assortative matching.
\newblock {\em American Economic Journal: Applied Economics}, 3(3):124--157.

\bibitem[Acemoglu et~al., 2004]{Acemoglu2004}
Acemoglu, D., Autor, D.~H., and Lyle, D. (2004).
\newblock Women, war, and wages: {T}he effect of female labor supply on the
  wage structure at midcentury.
\newblock {\em Journal of Political Economy}, 112(3):497--551.

\bibitem[Akbulut-Yuksel et~al., 2022]{AkbulutYuksel2022}
Akbulut-Yuksel, M., Tekin, E., and Turan, B. (2022).
\newblock {World War II} blues: {T}he long-lasting mental health effect of
  childhood trauma.
\newblock NBER Working Paper 30284.

\bibitem[Alwin et~al., 1992]{Alwin1992}
Alwin, D.~F., Braun, M., and Scott, J. (1992).
\newblock The separation of work and the family: {A}ttitudes towards women's
  labour-force participation in {G}ermany, {G}reat {B}ritain, and the {U}nited
  {S}tates.
\newblock {\em European Sociological Review}, 8(1):13--37.

\bibitem[Bastiaans, 2023]{Bastiaans2023}
Bastiaans, M. (2023).
\newblock Female labor supply and intergenerational spillovers: Evidence from a
  tax reform.
\newblock Working paper, {Erasmus University Rotterdam}.

\bibitem[Battistin et~al., 2022]{Battistin2022}
Battistin, E., Becker, S.~O., and Nunziata, L. (2022).
\newblock More choice for men? {M}arriage patterns after {W}orld {W}ar {II} in
  {I}taly.
\newblock {\em Journal of Demographic Economics}, 88(3):447--472.

\bibitem[Bellou and Cardia, 2016]{Bellou2016}
Bellou, A. and Cardia, E. (2016).
\newblock Occupations after {WWII}: {T}he legacy of {Rosie the Riveter}.
\newblock {\em Explorations in Economic History}, 62:124--142.

\bibitem[Bethmann and Kvasnicka, 2013]{Bethmann2013}
Bethmann, D. and Kvasnicka, M. (2013).
\newblock {World War II}, missing men and out of wedlock childbearing.
\newblock {\em The Economic Journal}, 123(567):162--194.

\bibitem[Bette, 2015]{Bette2015}
Bette, P. (2015).
\newblock War widows.
\newblock {\em 1914-1918-Online International Encyclopedia of the First World
  War}.

\bibitem[Boehnke and Gay, 2022]{Boehnke2022}
Boehnke, J. and Gay, V. (2022).
\newblock The missing men: {W}orld {W}ar {I} and female labor force
  participation.
\newblock {\em Journal of Human Resources}, 57(4):1209--1241.

\bibitem[Brainerd, 2017]{Brainerd2017}
Brainerd, E. (2017).
\newblock The lasting effect of sex ratio imbalance on marriage and family:
  Evidence from {W}orld {W}ar {II} in {R}ussia.
\newblock {\em The Review of Economics and Statistics}, 99(2):229--242.

\bibitem[Braun and Stuhler, 2023]{Braun2023}
Braun, S.~T. and Stuhler, J. (2023).
\newblock Exposure to war and its labor market consequences over the life
  cycle.
\newblock Kiel Working Paper 2241, Kiel Institute for the World Economy.

\bibitem[Brodeur and Kattan, 2022]{Brodeur2022}
Brodeur, A. and Kattan, L. (2022).
\newblock {World War II}, the baby boom, and employment: {C}ounty-level
  evidence.
\newblock {\em Journal of Labor Economics}, 40(2):437--471.

\bibitem[Broun{\'{e}}us et~al., 2024]{Brouneus2023}
Broun{\'{e}}us, K., Forsberg, E., H\"{o}glund, K., and Lonergan, K. (2024).
\newblock The burden of war widows: {G}endered consequences of war and
  peace-building in {S}ri {L}anka.
\newblock {\em Third World Quarterly}, 45(3):458--474.

\bibitem[Br\"{u}ck and Schindler, 2009]{Brueck2009}
Br\"{u}ck, T. and Schindler, K. (2009).
\newblock The impact of violent conflicts on households: What do we know and
  what should we know about war widows?
\newblock {\em Oxford Development Studies}, 37(3):289--309.

\bibitem[{Bundesamt f\"{u}r Justiz} and BMJV, 2020]{BundesamtJustiz2020}
{Bundesamt f\"{u}r Justiz} and BMJV (2020).
\newblock {Durchschnittliches Bruttoarbeitseinkommen der
  vollzeitbesch\"{a}ftigten Arbeitnehmer in der Bundesrepublik Deutschland in
  den Jahren 1949 bis 1989 (in Deutsche Mark)}.
\newblock Statista.

\bibitem[Burkhauser et~al., 2005]{Burkhauser2005}
Burkhauser, R.~V., Giles, P., Lillard, D.~R., and Schwarze, J. (2005).
\newblock Until death do us part: An analysis of the economic well-being of
  widows in four countries.
\newblock {\em The Journals of Gerontology Series B: Psychological Sciences and
  Social Sciences}, 60(5):S238--S246.

\bibitem[Buvinic et~al., 2012]{Buvinic2012}
Buvinic, M., Das~Gupta, M., Casabonne, U., and Verwimp, P. (2012).
\newblock Violent conflict and gender inequality: an overview.
\newblock {\em The World Bank Research Observer}, 28(1):110--138.

\bibitem[Buvinic et~al., 2013]{Buvinic2013}
Buvinic, M., Gupta, M.~D., Casabonne, U., and Verwimp, P. (2013).
\newblock Violent conflict and gender inequality: An overview.
\newblock {\em The World Bank Research Observer}, 28(1):110--138.

\bibitem[Chandran, 2020]{Chandran2020}
Chandran, R. (2020).
\newblock After four decades of war, {A}fghan widows battle for homes.
\newblock {\em Reuters}.
\newblock visited on 2022-12-14.

\bibitem[Diehl, 1985]{Diehl1985}
Diehl, J.~M. (1985).
\newblock Change and continuity in the treatment of {G}erman {K}riegsopfer.
\newblock {\em Central European History}, 18(2):170--187.

\bibitem[Doepke et~al., 2015]{Doepke2015}
Doepke, M., Hazan, M., and Maoz, Y.~D. (2015).
\newblock The baby boom and {World War II}: {A} macroeconomic analysis.
\newblock {\em The Review of Economic Studies}, 82(3):1031--1073.

\bibitem[Dr\`{e}ze and Srinivasan, 1997]{Dreze1997}
Dr\`{e}ze, J. and Srinivasan, P.~V. (1997).
\newblock Widowhood and poverty in rural {I}ndia: some inferences from
  household survey data.
\newblock {\em Journal of Development Economics}, 54(2):217--234.

\bibitem[Dupraz and Ferrara, 2023]{Dupraz2023}
Dupraz, Y. and Ferrara, A. (2023).
\newblock Fatherless: {T}he long-term effects of losing a father in the {U.S.
  Civil War}.
\newblock {\em Journal of Human Resources}, forthcoming.

\bibitem[Fadlon and Nielsen, 2021]{Fadlon2021}
Fadlon, I. and Nielsen, T.~H. (2021).
\newblock Family labor supply responses to severe health shocks: Evidence from
  {D}anish administrative records.
\newblock {\em American Economic Journal: Applied Economics}, 13(3):1--30.

\bibitem[Fern\'{a}ndez, 2013]{Fernandez2013}
Fern\'{a}ndez, R. (2013).
\newblock Cultural change as learning: {T}he evolution of female labor force
  participation over a century.
\newblock {\em American Economic Review}, 103(1):472--500.

\bibitem[Fern\'{a}ndez et~al., 2004]{Fernandez2004}
Fern\'{a}ndez, R., Fogli, A., and Olivetti, C. (2004).
\newblock Mothers and sons: {P}reference formation and female labor force
  dynamics.
\newblock {\em The Quarterly Journal of Economics}, 119(4):1249--1299.

\bibitem[Gay, 2023]{Gay2023}
Gay, V. (2023).
\newblock The intergenerational transmission of {World War I} on female labour.
\newblock {\em The Economic Journal}, 133(654):2303--2333.

\bibitem[{GESIS - Leibniz-Institut f{\"u}r Sozialwissenschaften},
  2002a]{ZA1670}
{GESIS - Leibniz-Institut f{\"u}r Sozialwissenschaften} (2002a).
\newblock {Allgemeine Bev{\"o}lkerungsumfrage der Sozialwissenschaften ALLBUS
  1988}.
\newblock GESIS Datenarchiv, K{\"o}ln. ZA1670 Datenfile Version 1.0.0,
  https://doi.org/10.4232/1.1670.

\bibitem[{GESIS - Leibniz-Institut f{\"u}r Sozialwissenschaften},
  2002b]{ZA1335}
{GESIS - Leibniz-Institut f{\"u}r Sozialwissenschaften} (2002b).
\newblock {Allgemeine Bev{\"o}lkerungsumfrage der Sozialwissenschaften ALLBUS
  Kumulierter Datensatz 1980-1986}.
\newblock GESIS Datenarchiv, K{\"o}ln. ZA1335 Datenfile Version 1.0.0,
  https://doi.org/10.4232/1.1335.

\bibitem[Goldin and Olivetti, 2013]{Goldin2013}
Goldin, C. and Olivetti, C. (2013).
\newblock Shocking labor supply: {A} reassessment of the role of {World War II}
  on women's labor supply.
\newblock {\em American Economic Review}, 103(3):257--262.

\bibitem[Goldin, 1991]{Goldin1991}
Goldin, C.~D. (1991).
\newblock The role of {World War II} in the rise of women's employment.
\newblock {\em American Economic Review}, 81(4):741--756.

\bibitem[Ichino and Winter-Ebmer, 2004]{Ichino2004}
Ichino, A. and Winter-Ebmer, R. (2004).
\newblock The long-run educational cost of {W}orld {W}ar {II}.
\newblock {\em Journal of Labor Economics}, 22(1):57--87.

\bibitem[Imbens and Wooldridge, 2009]{Imbens2009}
Imbens, G.~W. and Wooldridge, J.~M. (2009).
\newblock Recent developments in the econometrics of program evaluation.
\newblock {\em Journal of Economic Literature}, 47(1):5--86.

\bibitem[IRB, 2007]{CIRBC2007}
IRB (2007).
\newblock Afghanistan: Acquisition of widow status, both social and legal, when
  husband is missing or presumed dead; impact on a woman's status if a husband
  reappears after she has remarried; ability to divorce a first husband who has
  been missing for years (2003-2006).
\newblock AFG101116.E.

\bibitem[Jaworski, 2014]{Jaworski2014}
Jaworski, T. (2014).
\newblock ``{Y}ou're in the army now:'' {T}he impact of {World War II} on
  women's education, work, and family.
\newblock {\em The Journal of Economic History}, 74(1):169--195.

\bibitem[Kesternich et~al., 2020]{Kesternich2020}
Kesternich, I., Siflinger, B., Smith, J.~P., and Steckenleiter, C. (2020).
\newblock Unbalanced sex ratios in {G}ermany caused by {W}orld {W}ar {II} and
  their effect on fertility: A life cycle perspective.
\newblock {\em European Economic Review}, 130:103581.

\bibitem[Kesternich et~al., 2014]{Kesternich2014}
Kesternich, I., Siflinger, B., Smith, J.~P., and Winter, J.~K. (2014).
\newblock The effects of {W}orld {W}ar {II} on economic and health outcomes
  across {E}urope.
\newblock {\em The Review of Economics and Statistics}, 96(1):103--118.

\bibitem[Matysiak and Steinmetz, 2008]{Matysiak2008}
Matysiak, A. and Steinmetz, S. (2008).
\newblock Finding their way? female employment patterns in {West} {Germany},
  {East} {Germany}, and {Poland}.
\newblock {\em European Sociological Review}, 24(3):331--345.

\bibitem[Mayer, 1995]{ZA2647}
Mayer, K.~U. (1995).
\newblock {Lebensverl{\"a}ufe und gesellschaftlicher Wandel: Die
  Zwischenkriegskohorte im {\"U}bergang zum Ruhestand (Lebensverlaufsstudie
  LV-West II T - Telefonische Befragung)}.
\newblock GESIS Datenarchiv, K{\"o}ln. ZA2647 Datenfile Version 1.0.0,
  https://doi.org/10.4232/1.2647.

\bibitem[Mayer, 2018a]{ZA2646}
Mayer, K.~U. (2018a).
\newblock {Lebensverl{\"a}ufe und gesellschaftlicher Wandel: Die
  Zwischenkriegskohorte im {\"U}bergang zum Ruhestand (Lebensverlaufsstudie
  LV-West II A - Pers{\"o}nliche Befragung)}.
\newblock GESIS Datenarchiv, K{\"o}ln. ZA2646 Datenfile Version 1.1.0,
  https://doi.org/10.4232/1.13194.

\bibitem[Mayer, 2018b]{ZA2645}
Mayer, K.~U. (2018b).
\newblock {Lebensverl{\"a}ufe und gesellschaftlicher Wandel: Lebensverl{\"a}ufe
  und Wohlfahrtsentwicklung (Lebensverlaufsstudie LV-West I)}.
\newblock GESIS Datenarchiv, K{\"o}ln. ZA2645 Datenfile Version 1.1.0,
  https://doi.org/10.4232/1.13193.

\bibitem[Mednick, 2018]{Mednick2018}
Mednick, S. (2018).
\newblock War widows scrape by in south sudan's opposition stronghold.
\newblock {\em The New Humanitarian}.

\bibitem[Mikrozensus-Zusatzerhebung, 1971]{Gesis1971}
Mikrozensus-Zusatzerhebung (1971).
\newblock {\em {Berufliche und soziale Umschichtung der Bev\"{o}lkerung}}.
\newblock GESIS-Datenfile.

\bibitem[M\"{u}ller, 2016]{Mueller2016}
M\"{u}ller, R.-D. (2016).
\newblock {\em {Hitler's W}ehrmacht, 1935--1945}.
\newblock University Press of Kentucky.

\bibitem[Niehuss, 2002]{Niehuss2002}
Niehuss, M. (2002).
\newblock {\em {Familie, Frau und Gesellschaft. Studien zur Strukturgeschichte
  der Familie in Westdeutschland 1945-1960}}.
\newblock Vandenhoeck \& Ruprecht, G\"{o}ttingen.

\bibitem[Overmans, 1999]{Overmans1999}
Overmans, R. (1999).
\newblock {\em {Deutsche milit\"{a}rische Verluste im Zweiten Weltkrieg}}.
\newblock R. Oldenbourg, Munich.

\bibitem[Rose, 2018]{Rose2018}
Rose, E.~K. (2018).
\newblock The rise and fall of female labor force participation during {World
  War II} in the {United States}.
\newblock {\em The Journal of Economic History}, 78(3):673--711.

\bibitem[Salisbury, 2017]{Salisbury2017}
Salisbury, L. (2017).
\newblock Women{\textquotesingle}s income and marriage markets in the {U}nited
  {S}tates: {E}vidence from the {C}ivil {W}ar pension.
\newblock {\em The Journal of Economic History}, 77(1):1--38.

\bibitem[Schn\"{a}delbach, 2007]{Schnaedelbach2007}
Schn\"{a}delbach, A. (2007).
\newblock {\em {Kriegerwitwen. Lebensbew\"{a}ltigung zwischen Arbeit und
  Familie in Westdeutschland nach 1945}}.
\newblock Campus-Verlag, Frankfurt/New York.

\bibitem[Skocpol, 1993]{Skocpol1993}
Skocpol, T. (1993).
\newblock America's first social security system: {T}he expansion of benefits
  for civil war veterans.
\newblock {\em Political Science Quarterly}, 108(1):85--116.

\bibitem[{Statistisches Bundesamt}, 2023]{StatistischesBundesamt2023}
{Statistisches Bundesamt} (2023).
\newblock {\em Preise. {Verbraucherpreisindizes f\"{u}r Deutschland. Lange
  Reihen ab 1948}}.
\newblock {Statistisches Bundesamt}, Wiesbaden.

\bibitem[Tegtmeyer, 1979]{Tegtmeyer1979}
Tegtmeyer, H. (1979).
\newblock {\em {Soziale Strukturen und individuelle Mobilit\"{a}t. Beitr\"{a}ge
  zur sozio-demographischen Analyse der Bundesrepublik Deutschland}}, chapter
  Berufliche und soziale Umschichtung der Bev\"{o}lkerung: Methodische
  Anmerkungen zur Planung, Durchf\"{u}hrung und Aufbereitung der Befragung,
  pages 17--47.
\newblock Schriftenreihe des Bundesinstituts f\"{u}r Bev\"{o}lkerungsforschung.
  Boppard: Boldt.

\bibitem[{United Nations}, 2001]{UniNations2001}
{United Nations} (2001).
\newblock {\em Widowhood: {I}nvisible women, secluded or excluded}.
\newblock United Nations, New York.

\bibitem[York, 2023]{York2023}
York, J. (2023).
\newblock `too high a price': Ukraine's war widows forge a path towards an
  uncertain future.
\newblock {\em France24.com}.

\end{thebibliography}
	\end{small}
	\newpage
	
	\appendix
	\renewcommand{\thetable}{\Alph{table}}
	\renewcommand{\thetable}{\Alph{section}\arabic{table}}
	\renewcommand{\thefigure}{\Alph{section}\arabic{figure}}
	\renewcommand{\theequation}{\Alph{section}-\arabic{equation}}
	\setcounter{page}{1}
	\setcounter{equation}{0}

\section*{Online Appendix}

\section{Compensation for War Widows, 1951-1975}
\label{sec:compensation}

\setcounter{table}{0}
\setcounter{figure}{0}

\begin{figure}[!tbp]
	\caption{Maximal compensation for war widows, 1951-1975}\label{fig:pension}
	\centering
	\begin{threeparttable}
		\begin{center}
			\subfloat[Relative to gross labor income and real compensation index]{\includegraphics[clip, trim=0.5cm 0.7cm 0.5cm 0.7cm, width=0.75\textwidth]{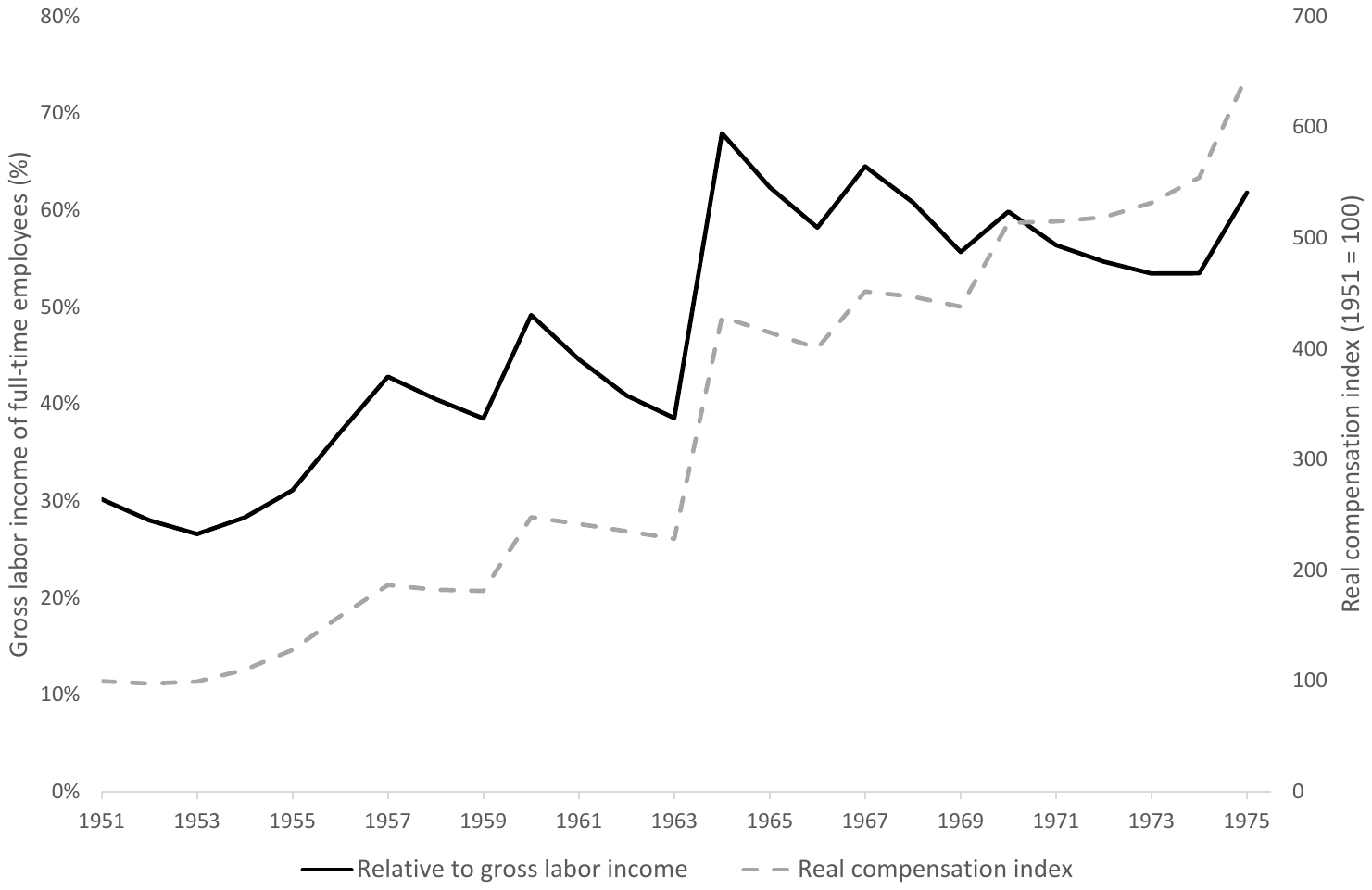}\label{fig:compensation}}
			\\
			\subfloat[Composition of compensation (relative to gross labor income)]{\includegraphics[clip, trim=0.5cm 0.7cm 0.5cm 0.7cm, width=0.75\textwidth]{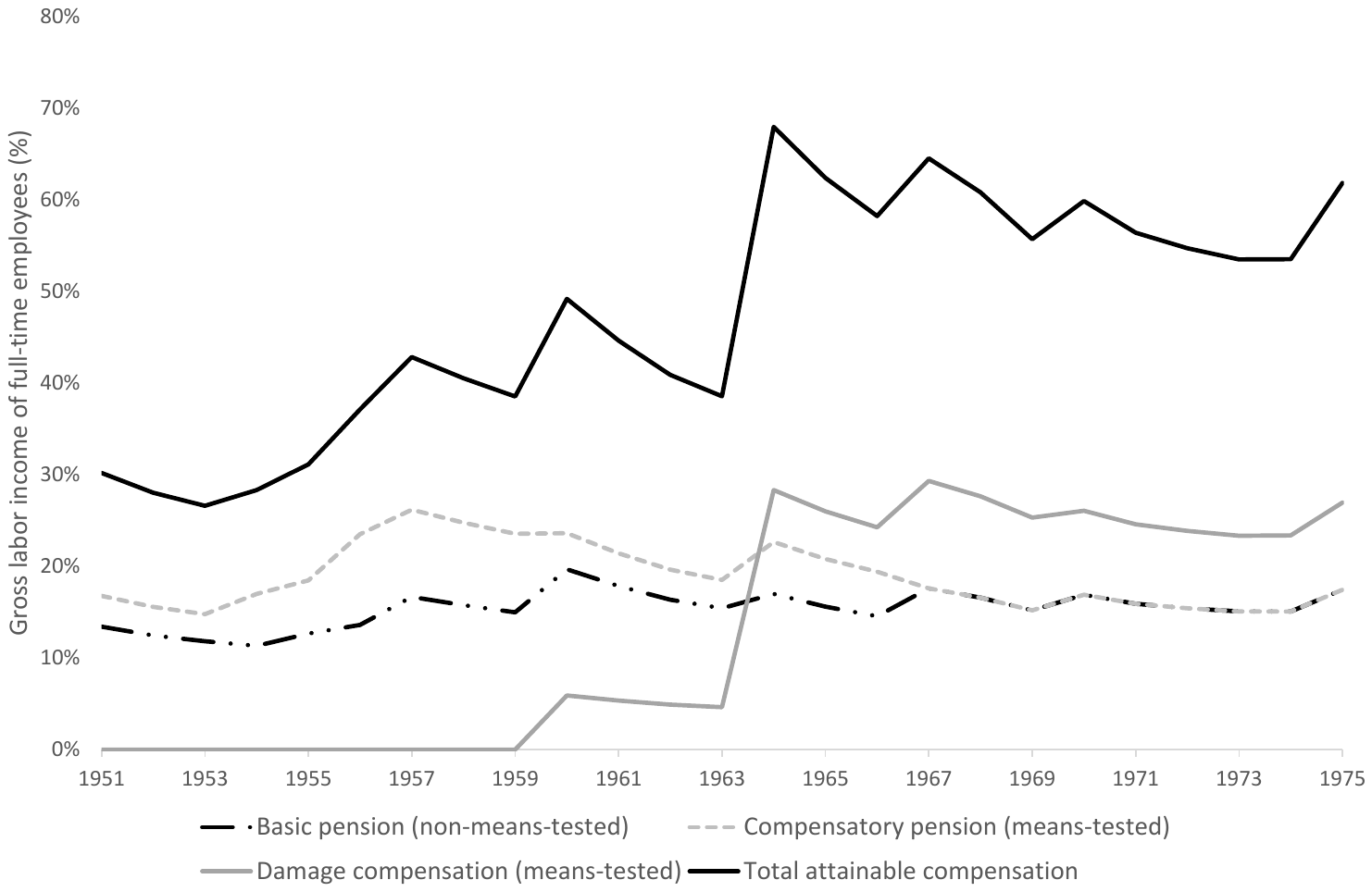}\label{fig:composition}}
		\end{center}
		\begin{tablenotes}
			\item \footnotesize{\emph{Notes}: The figure shows the maximum compensation for war widows between 1951 and 1975. Panel (a) shows total compensation relative to the gross labor income of full-time workers (black solid line) and the real increase in maximum compensation since 1951 (gray dashed line; the 1951 value is normalized to 100). Panel (b) decomposes total compensation relative to gross labor income (black solid line) into the part due to basic pension (non-means-tested; black dashed-dotted line), compensatory pension (means-tested; gray dashed line), and damage compensation (means-tested; gray solid line). See the description in Section \ref{Section Background} and Appendix \ref{sec:compensation} for further details.}
			\item \footnotesize{\textit{Source}: Author's calculations based on the \textit{Bundesversorgungsgesetz} (in its various versions). Data on average gross labor income are taken from \cite{BundesamtJustiz2020}; the price index (standard of living of a 4-person household with medium income) is taken from \cite{StatistischesBundesamt2023}.}
		\end{tablenotes}
	\end{threeparttable}
\end{figure}

Figure \ref{fig:pension} documents the evolution of the maximum attainable war widow's pension from 1951 to 1975. Panel \ref{fig:pension}(a) shows the increase in maximum compensation since the mid-1950s, both when measured relative to gross labor income (black solid line) and when considering the real increase since 1951 (gray dashed line). Relative to labor income, compensation nearly doubled from about 30\% in the early 1950s to between 50\% and 60\% after the second revision of the BVG in 1964.\footnote{The occasional declines are due to the fact that compensation payments were not increased in some years, while labor income generally rose rapidly during Germany's economic boom.}  At the same time, the maximum attainable war widow's pension increased fivefold in real terms between 1951 and 1970. The increase began in the second half of the 1950s. It continued relatively evenly after that, with a notable spike in 1964 when significant ``damage compensation'' (\textit{Schadensausgleich}) was introduced for widows whose income was less than half their deceased husband's expected income.

Figure \ref{fig:pension}(b) shows how the three different types of compensation--non-means-tested basic pension, means-tested compensatory pension, and means-tested damage compensation--have contributed to the increase in the ratio of war widows' pensions to average gross labor income. As can be seen, the basic pension was relatively stable over time, fluctuating between 13\% and 17\% of average labor income. The increase in the war widow's pension was initially largely due to the fact that the compensatory pension payments outpaced the growth in labor income. Later, the introduction of damage compensation spurred the increase in the maximum war widow's pension. By 1970, means-tested forms of compensation accounted for 72\% of the total maximum pension.  

\clearpage

\section{Alternative Definitions of the Control Group}
\label{sec:altcontrol}

	\setcounter{table}{0}
\setcounter{figure}{0}

Unfortunately, the MZU71 does not contain the respondents' entire marriage history, but only the year of the last marriage for those married in 1971. 
This poses two problems for the definition of our control group. First, we cannot exclude the possibility that divorced or widowed women in our control group married only after 1945. We consider this to be a minor problem because women in our sample were well above the average age at first marriage of 25.4 years in 1945. Second, married women who married after 1945, whom we exclude from the analysis, could in principle have been in an earlier marriage during the war (as we only observe their last year of marriage). Again, we consider this a minor problem, as less than 4\% of the married women in our sample married after 1945. Importantly, we show in the main text that our results hold in a second survey, the German Life History Study, where we observe women's complete marriage histories.

Nevertheless, in additional robustness checks, we drop widowed and divorced women from the control group and keep all married women in the control group. Table \ref{tab-lab-outcomes-rob} reports the results for our main results on labor force participation in 1950, 1960, and 1971. The results are virtually unchanged when we keep women who were married in 1971 in the control group, even if their last marriage was after 1945. This is to be expected since few women in our sample married after 1945. The differences are somewhat larger when we drop widowed and divorced women from the control group. This is to be expected since only married women remain in the control group--and these tend to have relatively low employment rates at the time. As a result, the initial employment gain from war widowhood is somewhat larger relative to baseline. But even compared to this control group, we find that war widows are more likely to be out of the labor force in 1971. However, this effect is entirely driven by the negative impact on the probability of being a helping family member. 

Table \ref{tab-lab-outcomes-rob} also shows that the counterintuitive effect of widowhood on labor force participation over the life cycle--elevated in 1950, but depressed in 1971--extends to younger cohorts born in 1915-21 and 1919-21. Almost all of these women were married to men born 1910-25 whose cohorts were fully conscripted during the war, so that selection into military service was negligible. %

\begin{table}[h!]
	\begin{center}
		\begin{threeparttable}
			\caption{Robustness: The impact of war widowhood on market work in 1950-1971 across samples} \centering
			\label{tab-lab-outcomes-rob}
			\begin{tiny}
				\begin{tabular}{lccccccccccc}
					\toprule\toprule
					&\multicolumn{3}{c}{1950} &&\multicolumn{3}{c}{1960}  &&\multicolumn{3}{c}{1971} \\ \cline{2-4} \cline{6-8} \cline{10-12}
					&Control  & 	&Obser-&& Control	&		& Obser-&& Control	& & Obser-\\
					&mean	  & IPW &vations&& Mean		& IPW	& vations && mean		& IPW & vations\\ 
					&  (1)	&(2)	& (3)	&& (4)	& (5) & (6) && (7) & (8) & (9) \\ \midrule
					\multicolumn{12}{l}{\textit{A. Market employment}:} \\				
					Baseline sample		& 0.205		& 0.138		& 29,302 && 0.238 & 0.058	& 30,342 && 0.162	& -0.019 & 30,351 \\
					& 			& (0.007)	&& 		 && (0.007)	&& 			&& (0.006)& \\
					Control group: Married women$^1$  & 0.156	& 0.176 	&21,106 && 0.174	& 0.102 & 21,860		&& 0.114 & 0.028 & 21,866 \\	
					\, (married 1945 or earlier)				  & 		& (0.007)	&&			&& (0.007)	&& 		 && (0.006) & \\
					Control group: Ever married  & 0.225	& 0.125	& 31,984 && 0.248	& 0.044		& 33,126&& 0.166 	& -0.017 & 33,137 \\
					women$^2$&				& (0.007) &&		&& (0.007)	&& 			&& (0.006) & \\ 

					Cohorts born 1915-21 & 0.227 	& 0.149	& 13,175	&& 0.293 & 0.048	&13,642 && 0.334	& -0.024 & 13,652 \\ 
					&			& (0.011) 	&&	 	& & (0.011)	&&		&	& (0.010)& \\
					Cohorts born 1919-21 & 0.241	& 0.127	& 6,185	&& 0.311 & 0.013	&6,414&& 0.366	&  -0.044 & 6,419 \\
					&			& (0.016)	&&		 && (0.016)  && 			&& (0.016) & \\ 
					\multicolumn{12}{l}{\textit{B. Out of the labor force}:} \\
					Baseline sample		&0.703 & -0.093	&29,302 && 0.677 & -0.003	&30,342&& 0.776	& 0.051 & 30,351\\
					&		& (0.007)	&& 		 && (0.007)	&&			&& (0.006) & \\
					Control group: Married women$^1$& 0.734	& -0.127 &21,106	&& 0.720	& -0.044  & 21,860 && 0.802 & 0.020 & 21,866 \\	
					\, (married 1945 or earlier) & 		& (0.008)	&&			&& (0.008)	&& 		& & (0.006) &\\
					Control group: Ever married 	& 0.680	& -0.079	&31,984 && 0.665	& 0.001	& 33,126	&& 0.769 	& 0.051 & 33,137  \\
					women$^2$ &		& (0.007) 	&&			&& (0.007)	&& 			&& (0.006) & \\							
					Cohort: Born 1915-21 & 0.696 & -0.127		&13,175 && 0.631 & -0.018	&13,642 && 0.604	& 0.048 & 13,652 \\ 
					&		 & (0.011) 		&&	 	& & (0.011)	&&			&& (0.011) & \\
					Cohort: Born 1919-21 & 0.684	& -0.109		&6,185 && 0.616 & 0.015	& 6,414 && 0.571	&  0.064 & 6,419 \\
					&			& (0.016)	&&		 && (0.016)  && 			&& (0.016) & \\ 					
					
						\bottomrule\bottomrule   
				\end{tabular}%
			\end{tiny}
			\begin{tablenotes}
				\item \tiny{\emph{Notes}: Means of the control group and IPW estimates for war widowhood across different samples of the 1971 Microcensus. The samples differ by the birth years of cohorts considered and the definition of the control group. Each estimate stems from a separate regression. Regression include as controls a full set of age dummies, an indicator for house ownership in 1939, indicators for expellees from Eastern Europe and refugees from GDR, the number of siblings, a full set of education dummies, indicators for the sector of employment in 1939 (agriculture, industry, construction, trade/transport, finance, public and private services, unknown) and the occupational or employment status in 1939 (self-employed, farmer, civil servant, white-collar worker, blue-collar worker, helping family member, out of the labor force including apprentices, in education, and unemployed). Robust standard errors clustered at selection districts (\textit{Auswahlbezirke}) are reported in parentheses. $^1$The control group consists only of women who were married in 1971 and whose marriage year was 1945 or earlier. Compared to the baseline sample, widowed and divorced women in the control group (as of 1971) are dropped. $^2$The control group includes all women who were ever married. Compared to the baseline sample, the control group also includes women who were married in 1971 but whose last marriage was after 1945.}
			\end{tablenotes}
		\end{threeparttable}
	\end{center}	
\end{table}

\section{Effect Heterogeneity}
\label{subsec:het}

\setcounter{table}{0}
\setcounter{figure}{0}

The main result of Section \ref{sec:results} is that war widowhood increased the probability of market employment only immediately after the war. In the longer run, war widowhood actually decreased market employment. Here, we document that the negative long-term effect on participation is strongest for less educated women with children.

Table \ref{tab-heterogeneity} presents estimates of the effect of widowhood on market employment in 1950, 1960, and 1971 separately for the subgroups indicated on the left. In the first row, we replicate our baseline results for ease of comparison. We first distinguish between women who have children and those who do not. Not surprisingly, women without children have much higher employment rates in middle age. For example, the control mean in 1950 is twice as high for those without children as for those with children (35.9\% versus 18.3\%). However, the pattern of the widowhood effect over time--large and positive in 1950 and then declining--is similar for both groups. As we can see, the negative long-term effect is only visible for women with children. For them, war widowhood reduced market employment by 2.4 pp in 1971.

Second, we show that high-educated women were much more likely than low-educated women to take up market employment after the death of their spouse in WWII. Moreover, for them, the positive effect persists, albeit muted, until late in life. We find that for highly educated women, war widowhood increased the probability of market employment by 21.9, 13.7, and 4.8 pp in 1950, 1960, and 1971, respectively. In contrast, the effect sizes are 12.2, 2.9, and -3.3 percentage points for women with low education. This differential effect is consistent with previous findings for the US that WWII mobilization increased female labor supply only among highly educated women \citep{Goldin2013}.

Finally, Table \ref{tab-heterogeneity} also examines the effect of widowhood by occupational status in 1939, distinguishing between women who were in market employment, who worked as helping family members, or who were out of the labor force before the war.  There are notable differences among these groups in 1950 and 1960, but not in 1971. In 1950, the widowhood effect is largest for those who worked as family helpers in 1939, increasing their probability of market work in 1950 by 26.1 percentage points (or 300\% compared to the control group's probability of 8.8\%). The increase is also substantial for women who did not work before the war, and more modest for those who were already in the labor force in 1939. The widowhood effect declines between 1950 and 1960 for all three groups, and it disappears by 1960 for those in market employment in 1939. This group also experienced the largest negative effect in 1971, at 3.6 percentage points.        

\begin{table}[h!]
	\begin{threeparttable}
		\caption{Heterogeneity in the impact of war widowhood on market employment in 1950, 1960, 1971} \centering
		\label{tab-heterogeneity}
		\begin{scriptsize}
			\begin{tabular}{lccccccccccc}
				\toprule\toprule
				&\multicolumn{3}{c}{1950} &&\multicolumn{3}{c}{1960}  &&\multicolumn{3}{c}{1971} \\ \cline{2-4} \cline{6-8} \cline{10-12}
				&Control  & 	&Obser-&& Control	&& Observ-		&& Control	& & Obser- \\
				&mean	  & IPW &vations&& Mean		& IPW	&vations&& mean		& IPW & vations\\ 
				&  (1)	&(2)	& (3)	&& (4)	& (5) & (6) && (7) & (8) & (9) \\ \midrule
				Baseline		& 0.205	& 0.138	& 29,302 & & 0.238 & 0.058	&30,342 && 0.162	& -0.019 & 30,351 \\
				& 		& (0.007)	&& 		 && (0.007)	&& 			&& (0.006) & \\ 
				
				\multicolumn{9}{l}{\textit{Children}:} \\
				With kids		& 0.183			& 0.133		& 25,331 & & 0.222	& 0.041 & 26,453		&& 0.157 & -0.024 & 26,444\\
								&				& (0.008)	&&&			& (0.007)	&& 		&& (0.006) & \\
				Without kids	& 0.359			& 0.159		& 3,771&& 0.350	& 0.080	&3,907&& 0.198	& 0.009 & 3,907 \\
				&				& (0.020)		&&			&& (0.021)	& & 	&&(0.018) &\\ 
				\multicolumn{9}{l}{\textit{Education}:} \\
				High ($>$10 years)	& 0.283	& 0.219		&5,232 && 0.336 & 0.137	& 5,408 && 0.244	& 0.048 & 5,408 \\
				&		& (0.019)	&&		 && (0.019)	&&			&& (0.018)& \\
				Low ($\leq$10 years)& 0.188	& 0.122		& 24,070 && 0.217 & 0.029& 24,934	&& 0.144	& -0.033 & 24,943\\	
				&		& (0.008)	&& 		 && (0.008)	&&			&& (0.006) & \\ 
				\multicolumn{9}{l}{\textit{Occupational status 1939}:} \\
				Market employment & 0.461	& 0.104 & 8,627		&& 0.446	& -0.000	& 8,985 && 0.253	& -0.036  & 8,985\\
				&				& (0.014)	&	 &&			& (0.014)	&& & &(0.012)& \\
				Helping family	& 0.088			& 0.261	& 2,838	&& 0.142	& 0.125	&2,945	&& 0.136 & -0.008 & 2,945 \\
				&				& (0.025)	&&			&& (0.025)	&&		&& (0.019)& \\
				Out of the labor force & 0.104 	& 0.137	&	17,820 && 0.156	& 0.061 & 18,393		&& 0.123 & -0.012 & 18,401 \\
				&				& (0.009)	&& 			&& (0.009)	&& 		&& (0.007) &\\
				\bottomrule\bottomrule   
			\end{tabular}%
		\end{scriptsize}
		\begin{tablenotes}
			\item \scriptsize{\emph{Notes}: Means of the control group and IPW estimates for war widowhood. Each estimate stems from a separate regression for the subgroup indicated on the left. Regression include as controls a full set of age dummies, an indicator for house ownership in 1939, indicators for expellees from Eastern Europe and refugees from GDR, the number of siblings, a full set of education dummies, indicators for the sector of employment in 1939 (agriculture, industry, construction, trade/transport, finance, public and private services, unknown) and the occupational or employment status in 1939 (self-employed, farmer, civil servant, white-collar worker, blue-collar worker, helping family member, out of the labor force including apprentices, in education, and unemployed). Robust standard errors clustered at selection districts (\textit{Auswahlbezirke}) are reported in parentheses.}
		\end{tablenotes}
	\end{threeparttable}
\end{table}

\section{Evidence on Attitudes towards Work and Gender Norms}
\label{sec:norms}

\setcounter{table}{0}
\setcounter{figure}{0}

This section summarizes evidence of the impact of war widowhood on attitudes towards work and gender norms. We first describe the data source and then discuss findings separately for war widows and their children.

\subsection{Data Description}
The ALLBUS is a survey that has been collecting data on the attitudes, behavior, and social structure of the German population every two years since 1980. We use data from five waves: 1980, 1982, 1984, 1986, 1988 \citep{ZA1335,ZA1670}. Each wave interviewed a random sample of about 3000 West German citizens over the age of 18 (foreigners were excluded). The exact questions vary, so our outcome variables of interest are typically available only for a subset of waves.

All five waves contain information on the complete marital history of the respondents. This allows us to identify women who lost their spouse between 1939 and 1945 (our treatment group). We compare them with women who married before 1945 but did not lose their spouse (control group). We restrict the sample to women born in 1906-21 in order to look at similar cohorts as in the main analysis. Our regressions control for a full set of dummies for the year of birth and year of first marriage, year of interview, and the years of schooling of the respondent and her father. 

In addition, the 1988 wave includes the year of death of the respondent's father and mother. Thus, for this wave, we can also examine the children of war widows by comparing respondents whose fathers died in World War II and those whose fathers did not. We focus on cohorts born in 1929-45 and drop respondents whose mothers died before 1945. The regressions control for respondents' year of birth and their fathers' years of schooling. 

\subsection{Work Attitudes and Gender Norms of War Widows}

The ALLBUS waves of 1980, 1982, and 1986 asked respondents about the importance of different domains in life, including family, job, leisure, friends, relatives, religion, and politics, as measured on a scale from 1 (unimportant) to 7 (very important). Table \ref{tab:allbusg1difareas} shows that war widows placed a, on overage, 0.36 lower value on the importance of family and children in life (relative to a control mean of 6.40).\footnote{We find similar results for an alternative question that asked respondents in 1980, 1984, and 1984 whether they believed that one needs a family to be truly happy. War widows are 9.1 pp less likely to agree (relative to a baseline of 80.8\%).} %
For all other domains of life, the differences between war widows and the control group are small. In particular, we find no evidence that war widows attach greater importance to job and work in life. If anything, the impact is negative (but small and statistically insignificant).  
\begin{table}[!t]
	\begin{center}
		\begin{threeparttable}
			\def\sym#1{\ifmmode^{#1}\else\(^{#1}\)\fi}
			\caption{Impact of war widowhood on the importance of different life areas} \centering
			\label{tab:allbusg1difareas}
			\begin{footnotesize}	
				\begin{tabular}{lccccccc}
					\toprule 		
					& \multicolumn{7}{c}{Importance of different life areas (1-7):} \\ 	\cline{2-8}	
					& Family \&	& Job \&  	& Leisure \&	& 		 	& 			& Religion \& 	&  			\\ 
					& children 	& work		& recreation	& Friends	& Relatives	& church		& Politics \\ 
					\cline{2-8}			
					& (1) &(2) &(3)&(4) &(5) & (6) & (7)\\ 
					
					\midrule
					War widow   &      -0.364&      -0.083&      -0.029&       0.023&      -0.142&      -0.156&       0.035\\
					&     (0.143)&     (0.196)&     (0.144)&     (0.145)&     (0.158)&     (0.186)&     (0.173)\\
					\midrule
					Control mean&       6.404&       4.493&       5.110&       5.364&       5.323&       4.978&       3.749\\
					N           &         831&         832&         832&         831&         833&         833&         833\\
							
					\bottomrule
				\end{tabular}
			\end{footnotesize}
			\begin{tablenotes} \item \footnotesize{\emph{Notes}: Control means and estimates of the effect of war widowhood on the importance of different domains of life. The sample consists of women born in 1906-21 in the 1980, 1982, and 1986 ALLBUS waves. The outcome variables are measured on a scale from 1 (unimportant) to 7 (very important). Each estimate comes from a separate OLS regression, controlling for a full set of dummies for year of birth and year of first marriage, year of interview, and years of schooling of the respondent and her father. Robust standard errors are shown in parentheses.}
			\end{tablenotes}
		\end{threeparttable}
	\end{center}
\end{table}

The ALLBUS waves of 1982 and 1988 asked several questions about women's roles in family and work. Unfortunately, most questions changed between waves and are thus not directly comparable. We summarize respondents' attitudes towards gender roles by measuring the proportion of progressive statements with which respondents agreed.\footnote{We count the negation of traditional gender roles as a progressive statement.} Column (1) of table \ref{tab:allbusg1gender} shows that this ``progressive gender role index'' is 7.2 pp higher among war widows, albeit from a comparatively low baseline of 24.6\%. In other words, war widows tend to have more progressive gender roles but still disagree with most progressive statements (or agree with traditional statements).

\begin{table}[!t]
	\begin{center}
		\begin{threeparttable}
			\def\sym#1{\ifmmode^{#1}\else\(^{#1}\)\fi}
			\caption{Impact of war widowhood on work-related gender norms} \centering
			\label{tab:allbusg1gender}
			\begin{footnotesize}	
				\begin{tabular}{lccccc}
					\toprule 		
					& Progressive	& Do small kids	& \multicolumn{3}{c}{Should women not}	\\
					& gender  	& suffer if	& \multicolumn{3}{c}{work if they have}	\\ \cline{4-6}
					& roles index & mother works? 	& \,\, no kids? \,\,	& small kids?	& school  kids?  \\ 
					& (0-1)			& (0/1)				& (0/1)			& (0/1)			& (0/1)			\\
					\cline{2-6}			
					& (1) &(2) &(3)&(4) &(5) \\ 
					\midrule
					War widow   &       0.072&      -0.039&      -0.109&       0.027&       0.037\\
					&     (0.029)&     (0.040)&     (0.067)&     (0.054)&     (0.101)\\
					\midrule
					Control mean&       0.246&       0.897&       0.183&       0.878&       0.730 \\
					N           &         512&         503&         174&         176&         176 \\
					\midrule
					Waves & 82, 88 & 82, 88 & 88 & 88 & 88  \\					
					\bottomrule
				\end{tabular}
			\end{footnotesize}
			\begin{tablenotes} \item \footnotesize{\emph{Notes}:  Control means and estimates of the effect of war widowhood on work-related gender norms. The sample consists of women born in 1906-21 in the 1982 and 1988 ALLBUS waves. The outcome variable in Column (1) summarizes the responses to six and nine statements about work-related gender norms asked in the 1982 and 1988 waves, respectively. The indicator measures the proportion of progressive statements with which respondents agreed (including the negation of traditional statements). The outcome variables in Columns (2) through (5) are indicator variables indicating whether respondents agreed with the question in the table header. Each estimate comes from a separate OLS regression, controlling for a full set of dummies for year of birth and year of first marriage, year of interview, and years of schooling of the respondent and her father. Robust standard errors are shown in parentheses.}
			\end{tablenotes}
		\end{threeparttable}
	\end{center}
\end{table}

Of particular importance for women's labor supply decisions is the compatibility of work and child-care responsibilities. In West Germany, the birth of a child often led previously employed women to give up work altogether in favor of the family, or to interrupt their careers for many years \citep[e.g.][]{Matysiak2008}. Previous research based on the 1988 ALLBUS wave has shown that West Germans strongly disapproved of women working outside the home when there are preschool children at home, even more so than respondents in the UK and US \citep{Alwin1992}. 

The results in Columns (2) to (5) indicate that war widows were not an exception, as they were not more supportive of women with young children working. Column (2) shows that war widows were only slightly less likely to agree with the statement that young children suffer when the mother works, a statement with which nearly 90\% of control group respondents agreed. And while war widows were more likely to approve of women working in situations where childcare is not an issue (Column (3)), they were even slightly more likely to disapprove of women working when they have children (Columns (4) and (5)).

\subsection{Attitudes and Gender Norms of War Widows' Children}

Table \ref{tab:allbusg2gender} shows that also the children of war widows did not hold more progressive gender norms than their peers. Neither the progressive gender role index (Column (1)) nor agreement with the statement that young children suffer when mothers work (Column (2)) differs statistically significantly between the two groups. The same applies to the approval of women's work in Columns (3) to (5), with one exception: Daughters of war widows have a 13.8 pp higher probability of agreeing with the statement that women with young children should \textit{not} work (from a baseline probability of 78.5 pp). Thus, if anything, we find evidence that the experience of growing up without a father has made daughters of war widows less supportive of the compatibility of women's work and caring for young children, potentially because they experienced the challenges faced by their mothers in the postwar period. In any case, even among respondents born in 1929-45, most disapproved of women working outside the home when preschool children are at home.

\begin{table}[!t]
	\begin{center}
		\begin{threeparttable}
			\def\sym#1{\ifmmode^{#1}\else\(^{#1}\)\fi}
			\caption{Intergenerational spillovers on work-related gender norms} \centering
			\label{tab:allbusg2gender}
			\begin{footnotesize}	
				\begin{tabular}{lccccc}
					\toprule 		
					& Progressive	& Do small kids	& \multicolumn{3}{c}{Should women not}	 \\
					& gender  	& suffer if	& \multicolumn{3}{c}{work if they have}	\\ \cline{4-6} 
					& roles index & mother works? 	& \,\, no kids? \,\, & small kids?	& school kids? \\ 
					& (0-1)			& (0/1)				& (0/1)			& (0/1)			& (0/1)			 \\
					\cline{2-6}			
					& (1) &(2) &(3)&(4) &(5)  \\ 
					\midrule
\multicolumn{6}{c}{\textit{A. Daughters:}}\\
Father's death&       0.028&      -0.045&       0.007&       0.138&       0.001\\
&     (0.039)&     (0.079)&     (0.049)&     (0.072)&     (0.092)\\
\midrule
Control mean&       0.474&       0.730&       0.072&       0.785&       0.506\\
N           &         375&         361&         340&         345&         338\\ \midrule
\multicolumn{6}{c}{\textit{B. Sons:}}\\
Father's death&       0.017&       0.019&       0.090&      -0.048&       0.008\\
&     (0.046)&     (0.068)&     (0.068)&     (0.078)&     (0.088)\\
\midrule
Control mean&       0.434&       0.812&       0.072&       0.843&       0.570\\
N           &         312&         304&         287&         291&         288\\
					
					\bottomrule
				\end{tabular}
			\end{footnotesize}
			\begin{tablenotes} \item \footnotesize{\emph{Notes}: Control means and estimates of the effect of war widowhood on work-related gender norms of their children. The sample consists of respondents born in 1929-45 in the 1988 ALLBUS who did not lose their mother before 1945. Panel A. restricts the sample to women (daughters), Panel B. restricts the sample to men (sons). The outcome variable in Column (1) summarizes the responses to nine statements about work-related gender norms asked in the 1988 wave. The indicator measures the proportion of progressive statements with which respondents agreed (including the negation of traditional statements). The outcome variables in Columns (2) through (5) are indicator variables indicating whether respondents agreed with the question in the table header. Each estimate comes from a separate OLS regression, controlling for a respondents' year of birth and their fathers' years of schooling. Robust standard errors are shown in parentheses.}
			\end{tablenotes}
		\end{threeparttable}
	\end{center}
\end{table}

\clearpage

\section{Additional Tables and Figures}

\setcounter{table}{0}
\setcounter{figure}{0}

\begin{table}[!h]
	\begin{center}
		\begin{threeparttable}
			\caption{Predicting war widowhood status in the Microcensus 1971} \centering
			\label{tab:predwarwidow}
			\begin{footnotesize}	
				\begin{tabular}{lcccc}
					\toprule 		
					& \multicolumn{4}{c}{Dependent variable:} \\ 
					& \multicolumn{4}{c}{War widow (0/1)} \\ 
					\cline{2-5}			
					& (1) &(2) &(3)&(4) \\	\midrule				
					R2	&  	0.005 & 0.009 & 0.011 & 0.012 \\
					\midrule 
					Birth year & yes & yes & yes & yes \\
					Socio-demographic characteristics & no & yes & yes & yes \\
					Employment \& occupational status in 1939 & no & no & no & yes \\
					Sector of employment in 1939 & yes & yes & yes & yes \\
					\bottomrule
				\end{tabular}
			\end{footnotesize}
			\begin{tablenotes} \item \footnotesize{\emph{Notes}: The table reports the R2 from regressions relating war widowhood status in the German Microcensus 1971 to an increasing number of covariates. Regressions include the following prewar covariates: (1) full set of age dummies, (2) = (1) plus an indicator for house ownership in 1939, indicators for expellees from Eastern Europe and refugees from GDR, number of siblings, full set of education dummies, (3) = (2) plus eight categories for the occupational or employment status in 1939 (self-employed, farmer, civil servant, white-collar worker, blue-collar worker, apprentice, helping family member, out of the labor force, in education, and unemployed). (4) = (3) plus six categories for the sector of employment in 1939 (agriculture, industry, construction, trade/transport, finance, public and private services).}
			\end{tablenotes}
		\end{threeparttable}
	\end{center}
\end{table}

\begin{table}[!h]
	\begin{center}
		\begin{threeparttable}
			\caption{Exogeneity of war widowhood in GHS} \centering
			\label{tab:exoshocks}
			\begin{footnotesize}	
				\begin{tabular}{lccccc}
					\toprule 							
					&  mean	 & \multicolumn{4}{c}{Dependent variable: War widow (0/1)} \\ \cline{3-6}			
					&  (std. dev.)  & (1) &(2) &(3)&(4) \\
					\midrule	
					Birth year 			& 1920.06 & 0.008 & -0.006 & 0.012 & 0.018 \\
					& (0.81) & (0.021) & (0.021) & (0.021) & (0.024) \\
					\# siblings 		& 2.80 & -0.006 & -0.010* & -0.007 & -0.002 \\
					& (2.49) & (0.007) & (0.006) & (0.006) & (0.007) \\
					Years of schooling 	& 8.64 & -0.011 & -0.018 & -0.018 & -0.019 \\
					& (1.26) & (0.013) & (0.014) & (0.014) & (0.016) \\
					Spouse's schooling (years)	& 9.27 & & -0.004 & -0.001 & 0.000 \\
					& (1.93) & & (0.003) & (0.004) & (0.004) \\
					Spouse's birth year  & 1914.58 & & 0.004 & 0.007 & 0.008 \\
					& (4.07) & & (0.011) & (0.010) & (0.012) \\
					Marriage year 		& 1941.55 & & & -0.032*** & -0.038*** \\
					& (1.94) & & & (0.008) & (0.010) \\
					Father's schooling (years) & 8.62 & & & & -0.013 \\
					& (1.95) & & & & (0.009) \\
					Mother's schooling 	& 8.24 & & & & -0.014 \\
					& (0.94) & & & & (0.013) \\
					Father's occupational score & 40.82 & & & & 0.002 \\
					& (11.05) & & & & (0.002) \\
					R2 				& & 0.002 & 0.013 & 0.043 & 0.056  \\
					N 				& 523 & 521 & 411 & 411 & 340 \\		 	
					\bottomrule
				\end{tabular}
			\end{footnotesize}
			\begin{tablenotes} \item \footnotesize{\emph{Notes}: The table reports coefficient estimates from a regression of war widowhood status on a set of prewar individual, spousal, and parental characteristics for women born 1919-21. Robust standard errors in parentheses.}
			\end{tablenotes}
		\end{threeparttable}
	\end{center}
\end{table}

\begin{figure}[!h]
	\caption{Illness over the life-cycle}\label{fig:illness}
	\centering
	\begin{threeparttable}
		\begin{center}
			\includegraphics[clip, trim=0.1cm 0.1cm 0.1cm 0.1cm, width=0.7\textwidth]{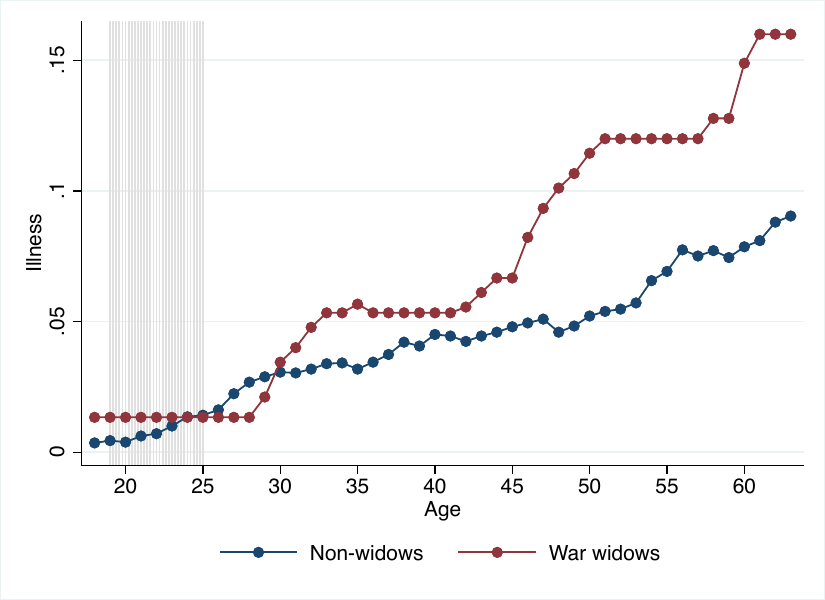}
		\end{center}
		\vspace*{-1mm}
		\begin{tablenotes}
			\item \footnotesize{\emph{Notes}: The figure shows the share of respondents reporting ill health over the life-cycle in the GHS, separately for war widows and non-widows with children born in or before 1945.}
		\end{tablenotes}
	\end{threeparttable}
\end{figure}

\begin{figure}[!h]
	\caption{Employment rate of war widows around the remarriage year}\label{fig:remarriage_eventstudy}
	\centering
	\begin{threeparttable}
		\begin{center}
			\includegraphics[clip, trim=0.1cm 0.1cm 0.1cm 0.1cm, width=0.7\textwidth]{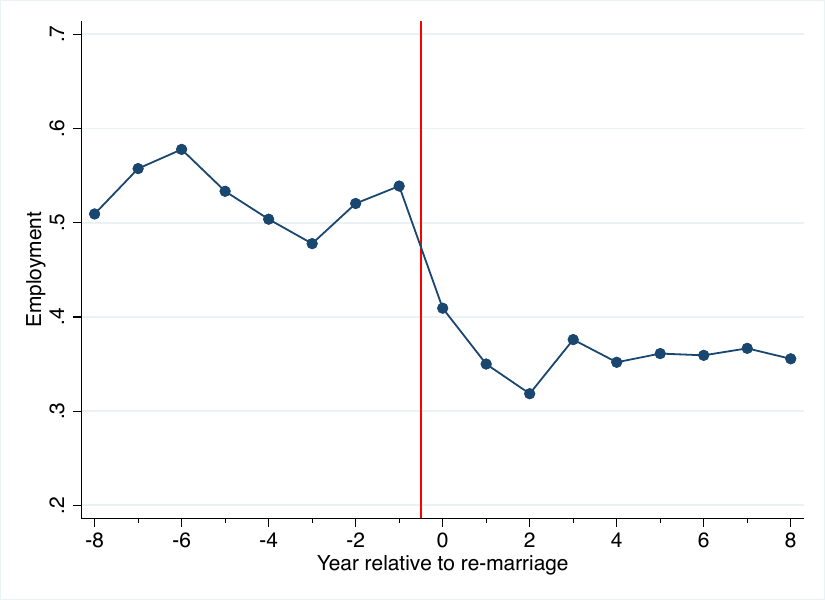}
		\end{center}
		\vspace*{-1mm}
		\begin{tablenotes}
			\item \footnotesize{\emph{Notes}: The figure shows the employment rate of war widows who remarried in the GHS. The x-axis is the event year relative to the year of remarriage.}
		\end{tablenotes}
	\end{threeparttable}
\end{figure}

\clearpage

\section{A Static Model of Labor Supply}
\label{sec:appendix_model}

\setcounter{table}{0}
\setcounter{figure}{0}

We consider a simple static model of labor supply in which mothers choose consumption $c$, hours of work $h$, leisure time $l$ and child-care time $t$ to maximize utility subject to a production function for child ``quality'' $q=q(t)$, with $q'>0$ and $q''<0$, and constraints for time $l_{0}=h+l+t$ and the budget, $c=wh+R_{0}$, where $w$ is the hourly wage rate and $R_{0}$ are other sources of household income (including labor income of the woman's spouse). 

Given the constraints for the budget and time, we can express consumption and hours worked as a function of leisure and child-care time (i.e., $h=l_{0}-l-t$), such that the individual's maximization problem simplifies to
\[
\underset{l,t}{max}~U(c(l,t),l,q(t);\theta).
\]
The parameter $\theta$ of the utility function represents the ``disutility of work'', which may vary with age or social norms, i.e. the ``stigma'' that society might place on working single mothers.\footnote{It might also vary with past work experience, either because of physical exhaustion or because working mothers accumulate experience that improves their attachment to the labor market.}

Taking first-order conditions with respect to $l$ and $t$, the optimal choices of leisure and child-care time are characterized by $\frac{U_l}{U_c}=w$, i.e., the marginal rate of substitution between consumption and leisure is equal to the wage, and $U_q/U_c=w/q'(t)$, i.e., the marginal rate of substitution between consumption and child quality is equal to the ratio between the wage and the marginal effect of child-care time on child quality. The optimal choices of hours worked and consumption follow from the time and budget constraints.

\begin{figure}[h]
	\caption{Labor supply of war widows}\label{fig:model}
	\centering
	\begin{threeparttable}
		\begin{center}
			\subfloat[Income shock from losing one's spouse]{\includegraphics[clip, width=0.43\textwidth]{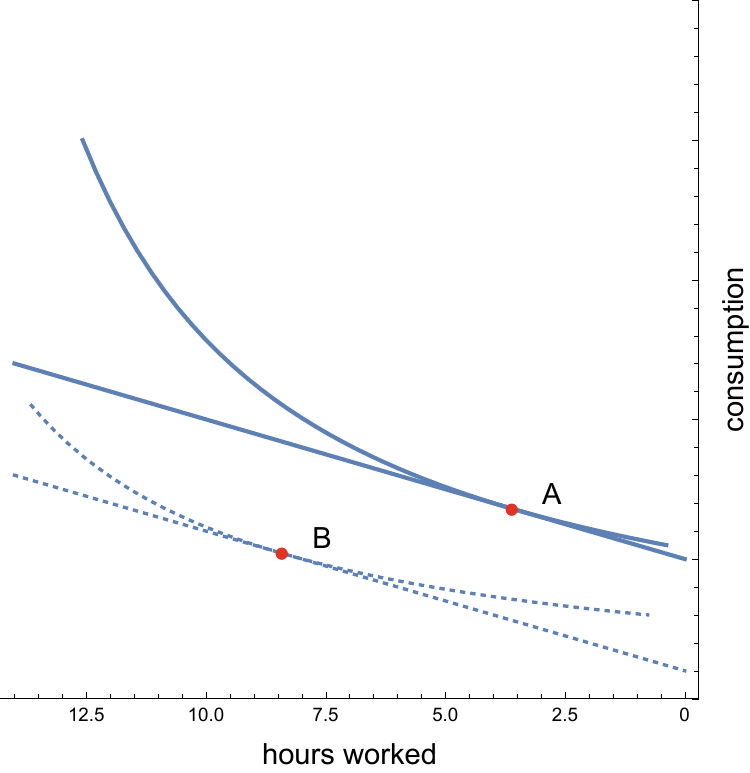}\label{fig:modela}}
			\, \, \, \, 
			\subfloat[Means-tested compensation]{\includegraphics[clip, width=0.43\textwidth]{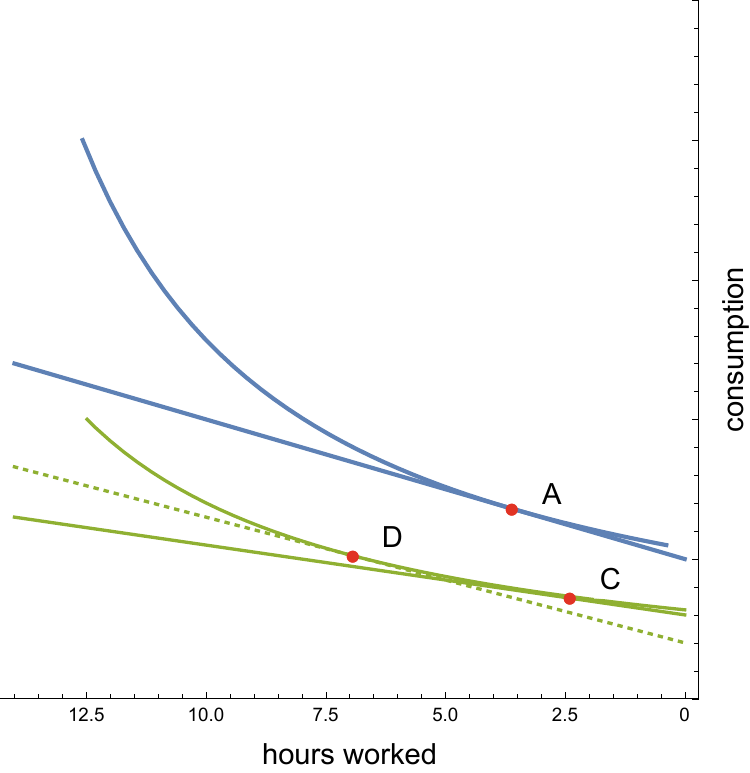}\label{fig:modelb}}
		\end{center}
		\vspace*{-1mm}
		\begin{tablenotes}
			\item \footnotesize{\emph{Notes}: Budget constraints and indifference curves in a static model of labor supply with $U=c^{1/2}l^{1/2}q(t)^{1/2}$ and $q(t)=t^{1/2}$. Sub-figure (a) illustrates the income effect from losing one's spouse, modeled as a reduction in the non-labor income $R_0$. Hours worked increase from point A to B. Sub-figure (b) illustrates the effect from a means-tested compensation scheme that partially offsets the reduction in $R_0$ but also reduces the effective take-home wage from work (decreasing hours worked from A to C, solid line), and an alternative policy in which compensation is less means-tested (increasing hours from A to D, dotted line).}
		\end{tablenotes}
	\end{threeparttable}
\end{figure}

Figure \ref{fig:model} provides an illustration how the loss of a spouse (sub-figure (a)) and means-tested compensation schemes (sub-figure (b)) affect hours worked $h$ and consumption $c$. For this illustration, we assume a Cobb-Douglas utility function with $U=l^{1/2}c^{1/2}q(t)^{1/2}$, $q(t)=t^{1/2}$ and no stigma. In sub-figure (a) we interpret the loss of a spouse as a negative income shock (i.e., a decrease in $R_{0}$).\footnote{Given the traditional gender norms at the time, the loss of a husband represents a greater shock to a household's income than time spent on childcare.} The budget curve decreases accordingly (dashed line). Assuming leisure is a normal good, such income loss decreases leisure, decreases time spent on childcare, increases participation, and increases working hours conditional on participation (\emph{income effect}). In the illustration, hours worked increase from point A to B. 

In sub-figure (b) we illustrate the effect of a compensation scheme that is partially means-tested. Because the income from additional hours of work crowds out compensation, the effective budget curve is flatter (green solid line) than the corresponding budget curve without compensation (blue line). This reduction in the effective wage rate disincentives labor supply (\emph{substitution effect}); in the example, compensation payments decrease hours worked from point A to C.

Sub-figure (b) also plots an alternative compensation policy that is characterized by lower unconditional payment (i.e., the budget curve has a higher intercept at $h=0$), which are however less rapidly withdrawn with labor income. Income therefore increases more rapidly with hours worked (dashed green line). As a consequence, the income effect dominates the substitution effect, and hours worked increase from point A to D.

\end{document}